\documentclass[letter,fleqn,usenatbib]{mnras}

\usepackage[T1]{fontenc}
\usepackage{ae,aecompl}
\usepackage{afterpage}

\usepackage{graphicx}	
\usepackage{amsmath}	
\usepackage{amssymb}	

\newcommand{\Msun}{${\rm M}_{\odot}$}

\newcommand{\Mstar}{$M_{*}$}

\newcommand{\Ha}{H$\alpha$}

\newcommand{\OII}{[O\,{\sc ii}\rm]}
\newcommand{\OIII}{[O\,{\sc iii}\rm]}
\newcommand{\sigoned}{$\sigma_{\rm 1D}$}
\newcommand{\sigintr}{$\sigma_{\rm intr}$}
\newcommand{\Rvir}{$R_{\rm vir}$}
\newcommand{\Minflow}{$\dot{M}_{\rm gas}$}

\newcommand\altaffilmark[1]{$^{#1}$}
\newcommand\altaffiltext[1]{$^{#1}$}

\title[FIRE kinematics]{What drives the evolution of gas kinematics in star-forming galaxies?}

\author[C.-L. Hung et al.]{
\parbox[t]{\textwidth}{ 
Chao-Ling Hung\thanks{E-mail: chaoling.hung@gmail.com}\altaffilmark{1},
Christopher C. Hayward\altaffilmark{2,3},
Tiantian Yuan\altaffilmark{4,5},
Michael Boylan-Kolchin\altaffilmark{6},
Claude-Andr{\'e} Faucher-Gigu{\`e}re\altaffilmark{7},
Philip Hopkins\altaffilmark{8},
Du\v{s}an Kere\v{s}\altaffilmark{9},
Norman Murray\altaffilmark{10},
Andrew Wetzel\altaffilmark{11}}
\vspace*{6pt} \\
\altaffiltext{1}{Physics Department, Manhattan College, 4513 Manhattan College Pkwy, Bronx, NY 10471, USA}\\
\altaffiltext{2}{Center for Computational Astrophysics, Flatiron Institute, 162 Fifth Avenue, New York, NY 10010, USA}\\
\altaffiltext{3}{Harvard-Smithsonian Center for Astrophysics, 60 Garden Street, Cambridge, MA 02138, USA}\\
\altaffiltext{4}{Centre for Astrophysics and Supercomputing, Swinburne University of Technology, Hawthorn, Victoria 3122, Australia}\\
\altaffiltext{5}{ARC Centre of Excellence for All Sky Astrophysics in 3 Dimensions (ASTRO 3D), Australia}\\
\altaffiltext{6}{Department of Astronomy, The University of Texas at Austin, Austin, TX 78712, USA}\\
\altaffiltext{7}{Department of Physics and Astronomy and CIERA, Northwestern University, 2145 Sheridan Road, Evanston, IL 60208, USA}\\
\altaffiltext{8}{TAPIR, Mailcode 350-17, California Institute of Technology, Pasadena, CA 91125, USA}\\
\altaffiltext{9}{Department of Physics, Center for Astrophysics and Space Sciences, University of California at San Diego, La Jolla, CA 92093}\\
\altaffiltext{10}{Canadian Institute for Theoretical Astrophysics, 60 St. George Street, University of Toronto, ON M5S 3H8, Canada}\\
\altaffiltext{11}{Department of Physics, University of California, Davis, CA 95616}
}

\date{Accepted XXX. Received YYY; in original form ZZZ}

\pubyear{2016}

\begin{document}
\label{firstpage}
\pagerange{\pageref{firstpage}--\pageref{lastpage}}
\maketitle

\begin{abstract}
One important result from recent large integral field spectrograph (IFS) surveys is that the intrinsic velocity dispersion of galaxies traced by star-forming gas increases with redshift.
Massive, rotation-dominated discs are already in place at $z\sim2$, but they are dynamically hotter than spiral galaxies in the local Universe. 
Although several plausible mechanisms for this elevated velocity dispersion (e.g. star formation feedback, elevated gas supply, or more frequent galaxy interactions) have been proposed, the fundamental driver of the velocity dispersion enhancement at high redshift remains unclear.
We investigate the origin of this kinematic evolution using a suite of cosmological simulations from the FIRE (Feedback In Realistic Environments) project.
Although IFS surveys generally cover a wider range of stellar masses than in these simulations, the simulated galaxies show trends between intrinsic velocity dispersion (\sigintr), SFR, and $z$ in agreement with observations.
In both the observed and simulated galaxies, \sigintr\ is positively correlated with SFR.
\sigintr\ increases with redshift out to $z\sim1$ and then flattens beyond that.
In the FIRE simulations, \sigintr\ can vary significantly on timescales of $\la100$ Myr.
These variations closely mirror the time evolution of the SFR and gas inflow rate (\Minflow). 
By cross-correlating pairs of \sigintr, \Minflow, and SFR, we show that increased gas inflow leads to subsequent enhanced star formation, and enhancements in \sigintr\ tend to temporally coincide with increases in \Minflow\ and SFR.
\end{abstract}




\section{Introduction}
The increasing capabilities of optical and near-infrared integral field spectrographs (IFSs) have revealed the internal dynamics of hundreds of star-forming galaxies out to $z\lesssim3$ \citep[see the review by][]{Glazebrook2013}. 
The majority of IFS surveys at $z>1$ probe the kinematics of ionised gas using tracers such as \Ha, \OIII, and \OII.
Whereas the spatially resolved kinematics traced by stars and other phases of gas have been routinely measured for nearby galaxies \citep[e.g.,][]{de-Zeeuw2002,Helfer2003,Walter2008,Falcon-Barroso2017}, at $z>1$, such measurements are either currently unattainable (stars and atomic gas) or only limited to a small number of galaxies \citep[molecular gas; e.g.,][]{Hodge2012}.
The IFS surveys at $1\lesssim z\lesssim3$ reveal that a significant fraction ($>1/3$) of massive galaxies (\Mstar\,$\gtrsim10^{10}$\Msun) exhibit smooth velocity gradients, indicative of rotating discs, whereas the rest exhibit irregular, merger-like kinematics or are dispersion-dominated \citep[e.g.][]{Flores2006,Forster-Schreiber2009,Law2009,Lemoine-Busserolle2010a,Gnerucci2011,Epinat2012,Wisnioski2015,Stott2016,Mieda2016,Mason2017}. 
Although the exact statistical breakdown depends on the sample selection, the spatial and spectral resolutions of the IFS observations, and the adopted galaxy classification schemes \citep[e.g.][]{Hung2015,Bellocchi2016,Rodrigues2016}, these results all point to an early emergence of massive, rotating disc-like galaxies at $z\gtrsim1$. 

A key difference between these massive rotating discs at $z\gtrsim1$ and spiral galaxies in the local Universe is that the intrinsic velocity dispersions traced by ionised gas (hereafter \sigintr) of $z\gtrsim1$ systems are significantly higher than those of their local counterparts (see \citealp{Yuan2017} for a rare exception).
These high$-z$ disc-like galaxies typically have \sigintr\ of $50-100$ km\,s$^{-1}$ and rotation velocity-to-dispersion ratios (V/\sigintr) of 1-5 \citep[][]{Law2009,Forster-Schreiber2009,Jones2010,Genzel2011}, although it is worth noting that the scatter within individual surveys is large and the measurement of \sigintr\ and beam-smearing corrections differ amongst surveys.
Nearby spiral galaxies tend to have a factor of 2-5 lower \sigintr\ \citep[e.g.][]{Epinat2010,Zhou2017}.
This trend of decreasing \sigintr\ with decreasing $z$ is also seen in slit-based observations \citep[e.g.][]{Kassin2012,Simons2016}.
In the local Universe, only more extreme systems -- such as ultraluminous and luminous infrared galaxies (ULIRGs and LIRGs), Lyman break galaxy (LBG)
analogues, and \Ha\ emitters -- exhibit \sigintr\ values comparable to those of high$-z$ systems \citep{Goncalves2010,Bellocchi2013,Green2014}.

The elevated \sigintr\ in high$-z$ galaxies has been associated with enhanced turbulent motions.
However, the physical driver(s) of the enhanced turbulence (e.g. feedback or gravitational instability) remains unclear.
\citet{Lehnert2009,Lehnert2013} show that there is a positive correlation between \sigintr\ and star formation rate (SFR) surface density at $z\sim2$ \citep[although see also][]{Genzel2011},
possibly suggesting that the enhanced \sigintr\ is driven by star formation feedback processes such as supernovae and radiation pressure \citep[e.g.][]{Thompson2005,Dib2006,
Ostriker2011,Shetty2012,Faucher-Giguere2013,Martizzi2015,Hayward2017,Orr2017}. Direct kinematic evidence for shells, bubbles, and outflows in M33 may also support such models \citep{Kam2015}.
Further evidence for a positive correlation between \sigintr\ and integrated SFR across a large redshift range is compiled in \citet{Green2014}.

It has also been claimed that gravitational instabilities due to external sources, such as cosmological gas accretion and galaxy interactions \citep[e.g.][]{Dekel2009a,Bournaud2011}, or internal dynamics, such as disc instabilities and clump-clump interactions \citep[e.g.][]{Wada2002,Agertz2009,Dekel2009a,Ceverino2010},
are responsible for the elevated \sigintr\ exhibited by high-$z$ galaxies.
Several works have found that the evolution of gas fraction (in a marginally stable disc) can explain the increase in \sigintr\ with $z$ \citep[e.g.][]{Swinbank2012a,Wisnioski2015,Turner2017}.
A recent analytic analysis by \citet{Krumholz2016} argues that the correlation between \sigintr\ and SFR is more consistent with gravitational instability-driven turbulence rather than stellar feedback-driven turbulence when galaxy gas fractions are taken into account.
However, this analytic work has yet to be tested with detailed simulations.

A number of observational effects can complicate the measurement of \sigintr\ and its physical interpretation.
It is increasingly difficult at higher redshift to probe the detailed kinematics of galaxies and make a fair comparison with their local counterparts due to surface brightness dimming and
limited spatial resolution.
Based on a set of artificially redshifted local spiral galaxies, \citet{Epinat2010} conclude that the mean kinematic properties can be recovered with proper disc modelling and beam smearing corrections.
However, this conclusion may not hold when galaxies exhibit more complicated intrinsic kinematic structures \citep[e.g.][]{Goncalves2010}.
Furthermore, to reduce a two-dimensional velocity dispersion map to a single value of \sigintr, some studies incorporate an isotropic velocity dispersion in the disc model and find a best-fitting \sigintr\ \citep[e.g.][]{Cresci2009}, whereas some studies simply employ a flux-weighted mean \citep[e.g.][]{Law2009}.
Some authors apply corrections for beam smearing \citep[e.g.][]{Stott2016}, whereas some studies do not; simulations suggest that the effects of beam smearing can be significant even for high-resolution \Ha\ kinematic maps of nearby galaxies \citep{Pineda2017}.

In this study, we aim to explore the physical drivers of \sigintr\ in star-forming galaxies using a suite of cosmological simulations from the FIRE project\footnote{\url{http://fire.northwestern.edu}} \citep[Feedback In Realistic Environments;][]{Hopkins2014}. 
In parallel, we gather measurements of \sigintr\ of star-forming galaxies at $0\leq z\lesssim3$ from IFS surveys presented in the literature.
We describe the simulation suite and the derivation of physical and kinematic properties in Section 2.
In Section 3, we compile a set of IFS observations from the literature and compare their physical and kinematic properties with the simulated galaxies.
We discuss possible physical drivers of the enhanced velocity dispersion of high-$z$ galaxies 
in Section 4.
Finally, we discuss the implications of our results in Section 5 and list our conclusions in Section 6.

\begin{figure*}
\centering
  \includegraphics[trim=0mm 0mm 20mm 0mm, clip=True, width=\textwidth]{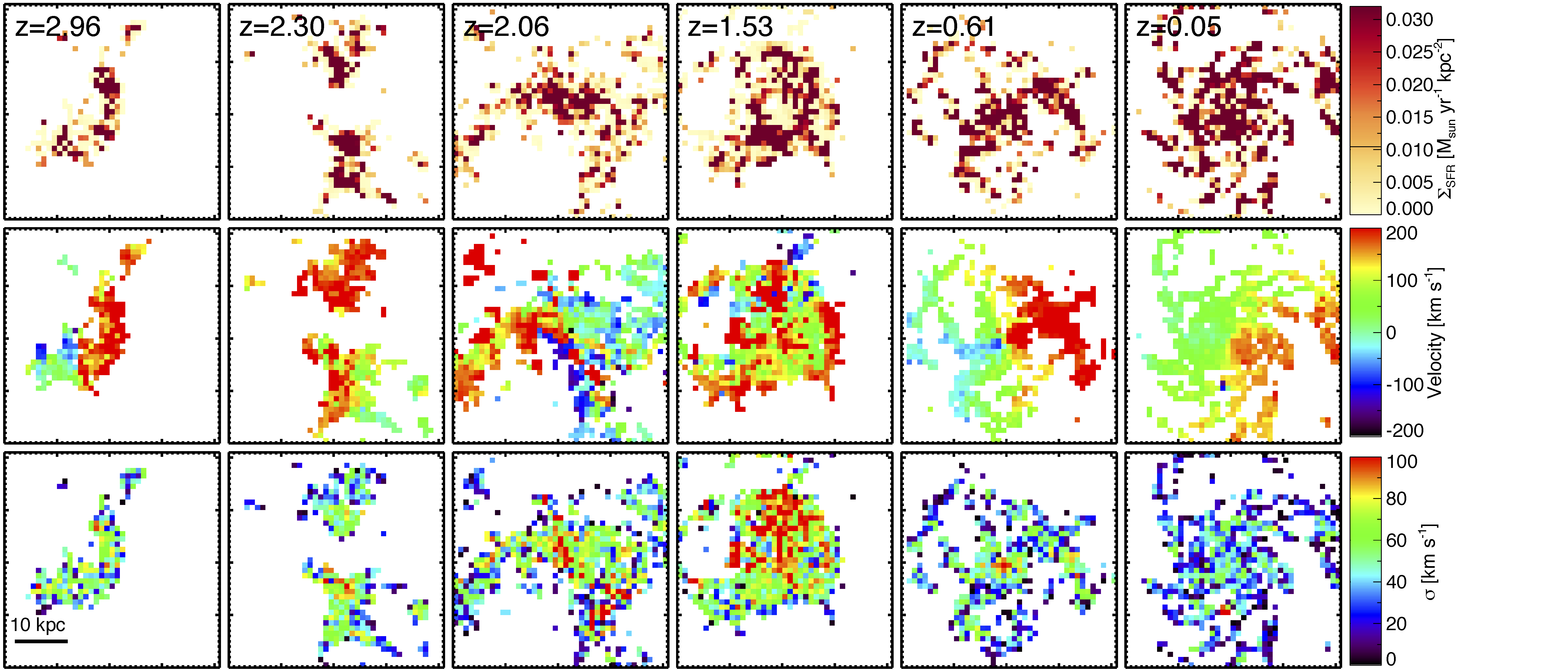} 
\caption{Selected snapshots from the m12i simulation, with each column corresponding to a different redshift
(indicated in the top-left corners of the top panels).
The top row shows SFR density (\Msun\,yr$^{-1}$\,kpc$^{-2}$) maps.
The middle and bottom rows show velocity and velocity dispersion maps, respectively, (in km\,s$^{-1}$) of the star-forming gas.
Each panel is 40 kpc $\times$ 40 kpc in size.
These maps are projected onto the x-y plane (arbitrarily defined in the simulations) for all redshift snapshots.
} 
\label{fig:m12v_stamp}
\end{figure*}

\section{Simulations and Analysis}

\subsection{FIRE simulations} \label{sec:fire_methods}

The simulations in this paper were run as part of the FIRE project; specifically, the original ``FIRE-1'' version of the code from \citet{Hopkins2014} was used.\footnote{We note
that after the bulk of the analysis in this paper was performed, some of the halos analysed in this work were re-simulated using the ``FIRE-2'' version of {\sc gizmo} \citep{Hopkins2017},
which employs an improved numerical method (the meshless finite mass scheme presented in \citealt{Hopkins2015gizmo}) and an improved algorithm for coupling
momentum and energy from stellar feedback to the ISM \citep{Hopkins2018SN}. To maintain consistency (not all of the halos included here have been re-simulated with
the FIRE-2 code) and avoid re-doing our analysis, we have opted to analyse the FIRE-1 simulations. However, we repeated our analysis for one of the FIRE-2 simulations (m12i)
and found that our conclusions are robust.}
They were run using the pressure-energy smoothed-particle hydrodynamics (``P-SPH'') mode of {\sc gizmo}\footnote{\url{http://www.tapir.caltech.edu/~phopkins/Site/GIZMO.html}} \citep{Hopkins2015gizmo}, a multi-method gravity plus hydrodynamics code. This formulation of SPH improves the treatment of fluid mixing instabilities and includes various other improvements to the artificial viscosity, artificial conductivity, higher-order kernels, and timestepping algorithm designed to reduce the most significant known discrepancies between SPH and grid methods \citep{Springel2010,Hopkins2013sph,Hayward2014arepo}. The gravity solver is an improved version of the Tree-PM solver from {\sc gadget-3} \citep{Springel2005gadget2}, with fully adaptive (and fully conservative) gravitational force softenings for gas following \citet{Price2007}.

The physics, source code, and all numerical parameters are described in detail in the papers above, but for completeness, we briefly review them here. Radiative heating and cooling is treated (via {\sc cloudy} tabulations; \citealt{Ferland1998}) from $10-10^{10}\,$K, including atomic, molecular, and metal-line cooling processes (following 11 species independently) and accounting for photo-heating both by a UV background \citep{FaucherGiguere2009} and local sources, in addition to self-shielding. Stars are spawned stochastically from gas that meets the following criteria: (1) it is self-gravitating according to the \citet{Hopkins2013sf_criteria} criterion, (2) it is molecular and self-shielding (following \citealt{Krumholz2011}), and (3) its density is greater than a minimum density threshold $n_{\rm min}\sim 5-50\,{\rm cm^{-3}}$, depending on the mass resolution -- and thus maximum density resolved -- of a given simulation.
The instantaneous star formation rate density is determined by dividing the molecular gas density by the local free-fall time (i.e. an instantaneous star formation efficiency of 100 per cent in the absence of feedback is assumed). 
However, stellar feedback disrupts clouds on timescales shorter than the local free-fall time, so the resulting star formation efficiency is less than 100\% except for at very high gas surface density \citep{Faucher-Giguere2013,Orr2017,Grudic2018}.

Once a star particle is formed, the simulations explicitly incorporate the following stellar feedback mechanisms: (1) local and long-range momentum flux from radiation pressure (both in the initial UV/optical single-scattering regime and re-radiated light in the IR); (2) energy, momentum, mass and metal injection from supernovae (Types Ia and II) and stellar mass loss (both OB and AGB stars); and (3) photo-ionisation and photo-electric heating. Every star particle is treated as a single-age stellar population with known mass, age, and metallicity. Given this information, all feedback event rates, luminosities, energies, mass-loss rates, and all other relevant quantities are tabulated directly from {\sc starburst99} \citep{Leitherer1999} stellar evolution models, assuming a \citet{Kroupa2001} IMF.
Note that AGN accretion and feedback are not implemented in these simulations.

The specific sample of simulations studied in this paper include simulations that were first presented in \citet{Hopkins2014} and \citet{FaucherGiguere2015}.
Specifically, we focus on three simulations (m12v, m12q, and m12i) from \citet{Hopkins2014}, for which the $z=0$ halo masses are in the range $\sim0.6-1.2\times10^{12}$ \Msun,
and the central galaxies have $z=0$ stellar masses of $\sim2.2-6.1\times10^{10}$ \Msun.
These haloes experience widely different accretion and merger histories: m12v, which uses higher-resolution initial conditions from \citet{Keres2009}, has a violent merger history with several encounters at $z<2$.
The m12q and m12i haloes, which were drawn from the AGORA project \citep{Kim2014}, have less violent histories than m12v.
Since the $z=2$ stellar mass values of the aforementioned simulations are about an order of magnitude lower than those of typical galaxies observed in high$-z$ IFS surveys (with the exception of lensed galaxies), we also analyse an additional eight simulations of more massive haloes from \citet{FaucherGiguere2015}, which were only run to $z=2$ (the ``z2h'' series).
The z2h simulations are representative of more-massive haloes at $z=2$: they have $z=2$ stellar masses of $0.3-4\times10^{10}$ \Msun\ and $z=2$ halo mass of $0.2-1.2\times10^{12}$ \Msun.

Whereas earlier studies have explored the role of accretion in driving turbulence via numerical experiments with idealised discs \citep[e.g.][]{Hopkins2013a} or analytic models \citep[e.g.][]{Elmegreen2010,Genel2012}, cosmological simulations are superior for studying this process because accretion of primordial gas at early times and recycled gas at later times is treated self-consistently and depends on e.g. environment in a manner that cannot be easily treated via idealised simulations and analytic models but is incorporated naturally in cosmological simulations.
Furthermore, these high-resolution simulations (with minimum baryonic force softening lengths of $\lesssim$10 pc) enable accurate treatments of stellar feedback processes, as summarised above, and thus allow us to study the effect of stellar feedback on the velocity dispersions of galactic gaseous discs.

\subsection{Physical properties of the FIRE galaxies}

We trace the evolution of the most massive halo in each simulation from $z=4$ to $z=0$ for the m12v, m12q, and m12i simulations and from $z=4$ to $z=2$ for the z2h simulations.
Using halo catalogs and merger trees generated by running Amiga's Halo Finder \citep[AHF;][]{Knollmann2009}, we determine the properties of the main halo and the central galaxy that it
hosts at each snapshot time.
We derive the kinematic properties of the simulated galaxies based on the dynamical information traced by the star-forming gas (i.e. SPH particles with SFR $>0$);
this is motivated by the goal to compare with recent IFS surveys that use nebular lines (e.g. \Ha, \OIII, and \OII) as kinematic tracers.
Following \citet{Hung2016}, we construct projected velocity and velocity dispersion maps along a given line-of-sight by measuring the SFR-weighted median and standard deviation within 500 pc $\times$ 500 pc pixels.
Figure~\ref{fig:m12v_stamp} shows example SFR surface density and kinematic maps of selected snapshots of the m12i simulation.

We measure a set of quantities that may be physical drivers of or are well correlated with galaxy kinematics, including the SFR \citep[e.g.][]{Lehnert2009,Green2010,Green2014}, \Mstar\ \citep[e.g.][]{Stott2016}, and gas fraction \citep[e.g.][]{Wisnioski2015,Krumholz2016}.
The SFRs (\Mstar\ values) of the selected massive haloes are defined as the total SFR (\Mstar) of gas (star) particles within 0.2 \Rvir\ of the halo centre.\footnote{The SFR
is the ``instantaneous'' SFR, which most closely corresponds to the \Ha\ luminosity and other short-timescale tracers (see e.g. \citealt{Hayward2014}, \citet{Sparre2017} and \citealt{Orr2017} for discussions) and will in general
differ from longer-timescale tracers, such as the UV and FIR luminosities, for the simulated galaxies at $z \ga 1$, where the star formation histories of even the relatively
massive simulated galaxies considered here are very bursty \citep{Sparre2017,Faucher-Giguere2018}.}
For the comparison with observations presented in Section 3, we use the time-averaged SFR calculated based on young stars formed within 10\,Myr at a given redshift.
The gas fraction ($f_{\rm gas}$) is defined as the ratio of the gas mass to the sum of the gas and stellar masses within 0.2 \Rvir, where the gas mass is computed from the subset of gas particles with $n\geq1\,{\rm cm^{-3}}$.
SFR, \Mstar, and $f_{\rm gas}$ are insensitive to the chosen outer radius when a value of $\gtrsim0.1$ \Rvir\ is used.  
Although currently poorly constrained in observations, another quantity of interest is the gas inflow rate because gas inflow can lead to subsequent changes in the gas fraction \citep[e.g.][]{Tacconi2010} and/or contribute to the bulk and turbulent motion, especially in the case of a galaxy merger \citep[e.g.][]{Hung2016}.
Following \citet{Faucher-Giguere2011} and \citet{Muratov2015}, we measure the gas inflow rate as the instantaneous mass flux through a thin spherical shell with outer radius 0.3 \Rvir\ and inner radius 0.2 \Rvir:
\begin{equation}
\dot{M}_{\rm gas} = \sum_i \mathbf{v_{i}} \cdot \frac{\mathbf{r_{i}}}{|r_{i}|} m_{\rm gas, i}/dL,
\end{equation}
where $dL = 0.1$ \Rvir\ and the infalling gas is defined as gas particles with $\mathbf{v_{i}} \cdot \frac{\mathbf{r_{i}}}{|r_{i}|}<0$.

\begin{table*}
\centering
 \caption{Properties of simulated galaxies in each redshift bin}
 \label{tab:propertysim}
 \begin{tabular}{ccccccc}
 \hline
 \hline
 Redshift Bin &  Simulations & \Mstar & SFR$^{a}$ & \sigoned$^{b}$  & $\sigma_{\rm 1D, median}^{c}$ & Number$^{d}$\\
            & & ($10^{10}$\Msun) & (\Msun\,yr$^{-1}$) &  (km\,s$^{-1}$)  & (km\,s$^{-1}$)  &  \\
 \hline
$0\leq z<0.25$ & m &2.58$^{e\,+0.82}_{-0.84}$ & 4.8$^{+2.0}_{-3.3}$ & 16.0$^{+4.7}_{-4.9}$ &  119.0$^{+10.0}_{-8.4}$ & 95 \\
$0.25\leq z<0.5$ & m &1.71$^{+0.39}_{-0.40}$ & 4.3$^{+1.3}_{-1.4}$ & 25.0$^{+7.2}_{-8.0}$ &  125.5$^{+11.5}_{-11.2}$ & 100 \\
$0.5\leq z<1.0$ & m & 1.17$^{+0.28}_{-0.24}$ & 5.0$^{+1.3}_{-1.3}$ & 34.8$^{+12.1}_{-11.3}$ &  121.2$^{+15.7}_{-20.9}$ & 43 \\
$1.0\leq z<1.5$ & m & 0.71$^{+0.11}_{-0.11}$ & 6.7$^{+3.0}_{-3.8}$ & 50.1$^{+11.7}_{-13.0}$ &  71.3$^{+20.1}_{-20.7}$ & 19  \\
$1.5\leq z<2.0$ & m & 0.31$^{+0.15}_{-0.15}$ & 4.4$^{+3.2}_{-2.5}$ & 39.2$^{+8.8}_{-10.0}$ &  67.0$^{+18.4}_{-18.1}$ & 35 \\                    
$2.0\leq z<2.5$ & all & 0.29$^{+0.14}_{-0.12}$ & 5.9$^{+3.6}_{-3.6}$ & 44.0$^{+16.1}_{-18.3}$ &  87.4$^{+26.0}_{-28.5}$  & 59 \\
						  & m & 0.14$^{+0.05}_{-0.07}$ & 2.9$^{+1.1}_{-1.8}$ & 35.8$^{+1.0}_{-6.3}$ &  58.9$^{+11.6}_{-15.3}$ & 10 \\
                         & z2h & 0.31$^{+0.18}_{-0.17}$ & 6.3$^{+3.6}_{-3.6}$ & 47.3$^{+20.1}_{-18.0}$ &  90.5$^{+28.6}_{-27.2}$ & 49 \\
$2.5\leq z<3.0$ & all &0.20$^{+0.10}_{-0.14}$ & 7.1$^{+5.5}_{-5.8}$ & 46.0$^{+13.8}_{-15.8}$ & 84.3$^{+28.8}_{-27.9}$ &44 \\
                         & m & 0.10$^{+0.01}_{-0.07}$ & 4.6$^{+2.1}_{-3.4}$ & 47.2$^{+6.0}_{-4.9}$ &  72.4$^{+16.0}_{-7.9}$ & 11 \\
                         & z2h & 0.24$^{+0.06}_{-0.13}$ & 7.9$^{+5.8}_{-5.9}$ & 43.8$^{+14.2}_{-15.3}$ &  87.0$^{+26.1}_{-25.6}$ & 33 \\
$3.0\leq z<3.5$ & all &0.08$^{+0.03}_{-0.07}$ & 3.1$^{+3.0}_{-2.4}$ & 40.4$^{+16.3}_{-15.9}$ & 73.9$^{+24.2}_{-25.8}$ & 43 \\
                        & m & 0.03$^{+0.01}_{-0.01}$ & 1.3$^{+0.2}_{-0.5}$ & 31.1$^{+2.6}_{-5.8}$ &  53.3$^{+5.9}_{-5.2}$ & 12 \\
                         & z2h & 0.10$^{+0.07}_{-0.05}$ & 4.2$^{+4.8}_{-2.8}$ & 43.6$^{+20.0}_{-19.1}$ &  86.1$^{+34.7}_{-33.6}$ & 31 \\
 \hline
 \multicolumn{7}{l}{$^a$ Time-averaged SFR based on stars formed within the past 10\,Myr.}\\
 \multicolumn{7}{l}{$^b$ \sigoned\ is defined as the minimum value across $10^4$ viewing angles.}\\
 \multicolumn{7}{l}{$^c$ $\sigma_{\rm 1D, median}$ is the median 1D velocity dispersion from all viewing angles. These values are}\\
 \multicolumn{7}{l}{~~~significantly larger than \sigoned\, as they certainly include large scale motions, such as rotation.}\\
 \multicolumn{7}{l}{$^d$ Number of galaxy snapshots in each redshift bin with valid velocity dispersion measurements.}\\
 \multicolumn{7}{l}{$^e$ The median value in each redshift bin; the quoted errors represent $16-84$}\\
 \multicolumn{7}{l}{~~~per cent of the data in each redshift bin.}\\
  \end{tabular}
\end{table*}

 \begin{table*}
\centering
 \caption{Properties of observed galaxies}
 \label{tab:propertyobs}
 \begin{tabular}{@{}lcccccl}
 \hline
 \hline
 Survey & Number$^{a}$ &Redshift$^{b}$ & Median \Mstar & Median SFR & Median $\sigma$& References\\
            &        &      &($10^{10}$\Msun) & (\Msun\,yr$^{-1}$) & (km\,s$^{-1}$) \\
 \hline
GHASP    & 137 & 0.00508$^{+0.00346}_{-0.00366}$ & 1.10$^{c +0.99}_{-1.07}$  & 0.15$^{c +0.28}_{-0.15}$ & 24$^{+5}_{-5}$ & \citet{Epinat2008,Epinat2010} \\   
DYNAMO & 67 & 0.07722$^{+0.05472}_{-0.02125}$ & 1.62$^{+1.40}_{-1.45}$ & 8.3$^{+7.6}_{-8.1}$ & 41$^{+15}_{-17}$ & \citet{Green2014} \\
LBA        & 16 & 0.18$^{+0.02}_{-0.04}$ & 0.63$^{+0.37}_{-0.38}$ & 19.4$^{+0.6}_{-16.9}$ & 67$^{+7}_{-17}$ & \citet{Goncalves2010} \\
MUSE/KMOS        & 179 & 0.81$^{+0.32}_{-0.32}$ & 0.24$^{+0.63}_{-0.24}$ & ... & 32$^{+9}_{-9}$ & \citet{Swinbank2017} \\
KROSS       & 472 & 0.84$^{+0.06}_{-0.05}$ & 0.98$^{+0.76}_{-0.78}$ & 7.0$^{+4.1}_{-4.2}$ & 43$^{+18}_{-18}$ & \citet{Harrison2017} \\
   &  &  &  &  &  & \citet{Johnson2017}\\
IROCKS   & 23 &0.94$^{+0.09}_{-0.09}$ & 3.16$^{+1.85}_{-2.37}$ & 11.1$^{+9.4}_{-8.4}$ & 62$^{+5}_{-7}$ & \citet{Mieda2016}\\
MASSIV   & 48 &1.23$^{+0.17}_{-0.21}$ & 1.51$^{+2.20}_{-1.06}$ & 45.9$^{+25.3}_{-35.9}$ & 52$^{+23}_{-20}$ & \citet{Epinat2012};\\
   &  &  &  &  &  & \citet{Queyrel2012}\\
WiggleZ  & 13 &1.31$^{+0.02}_{-0.02}$ & 1.99$^{+3.02}_{-1.36}$ & 29.9$^{+10.7}_{-9.7}$ & 93$^{+9}_{-8}$ & \citet{Wisnioski2011} \\
KLASS & 32 & 1.41$^{+0.36}_{-0.48}$ & 0.54$^{+0.22}_{-0.52}$   & ...    &   54$^{+32}_{-28}$ &  \citet{Mason2017}\\
SHiZELS & 19 & ...$^{d}$ & 1.07$^{+0.92}_{-0.81}$ & 7$^{+3}_{-5}$ & 72$^{+19}_{-15}$ & \citet{Swinbank2012}; \\ 
   &  &  &  &  &  & \citet{Molina2016}\\
CASSOWARY & 11 & 2.13$^{+0.41}_{-0.10}$ & ...  & 36$^{+23}_{-20}$   &  61$^{+17}_{-16}$   & \citet{Leethochawalit2016} \\
SINS disc-like  & 12  & 2.21$^{+0.18}_{-0.17}$ & 3.80$^{+2.80}_{-3.18}$ & 100$^{+46}_{-84}$ & 52$^{+6}_{-17}$ & \citet{Cresci2009} \\
Lensed  & 17 & 2.21$^{+0.44}_{-0.72}$ & 0.32$^{+0.68}_{-0.25}$ & 5$^{+6}_{-4}$ & 60$^{+10}_{-10}$ &  \citet{Livermore2015} \\
Law OSIRIS  & 16 & 2.29$^{+0.13}_{-0.14}$ & 2.88$^{+0.21}_{-2.58}$ & 19$^{+9}_{-6}$ & 69$^{+17}_{-13}$ & \citet{Law2009} \\
KDS  & 32 & 3.37$^{+0.23}_{-0.27}$ & 0.50$^{+0.29}_{-0.40}$ & ... & 71$^{+22}_{-20}$ & \citet{Turner2017} \\
VVDS  & 3 &3.28, 3.28, 3.70 & 1.23, 1.62, 1.51 & 125, 427, 1257 & 72, 60, 78 & \citet{Lemoine-Busserolle2010a} \\
\hline
\multicolumn{7}{l}{$^a$ Number of galaxies in each survey with valid velocity dispersion measurements.}\\
\multicolumn{7}{l}{$^b$ The errors in redshift, \Mstar, SFR, and \sigoned\ represent $16-84$ per cent of the data in each survey.}\\
\multicolumn{7}{l}{$^c$ A subset of GHASP galaxies have \Mstar\ and SFR estimates from the MPA-JHU measurements based on SDSS DR8 spectra}\\
\multicolumn{7}{l}{\ \ \citep{Kauffmann2003,Brinchmann2004,Salim2007}.}\\
\multicolumn{7}{l}{$^d$ SHiZELS is comprised of galaxies in three redshift bins: $z\sim0.8$, 1.47, and 2.23.}\\
 \end{tabular}
\end{table*}

\begin{figure}
\centering
  \includegraphics[width=0.5\textwidth]{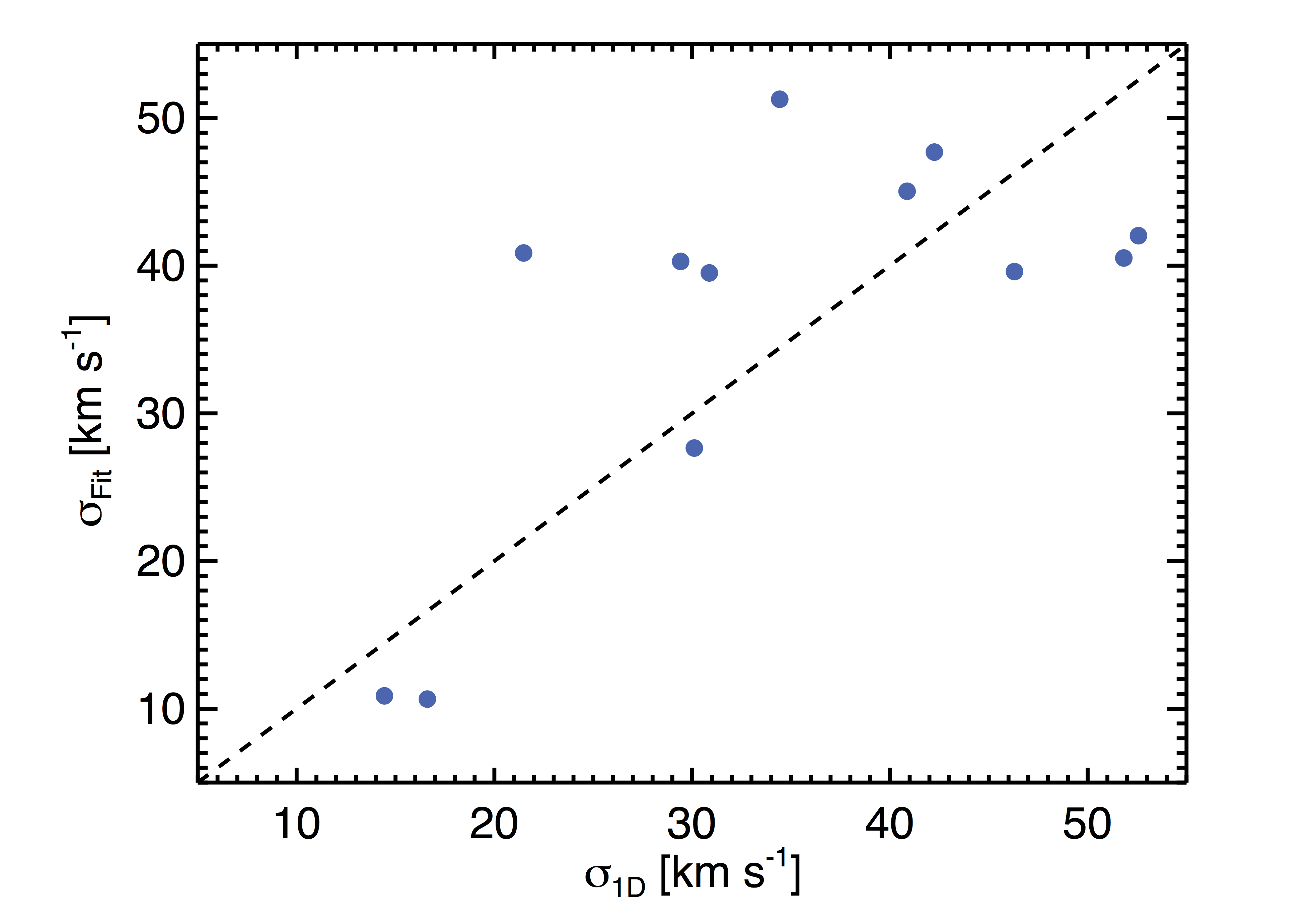} 
\caption{\sigoned\ (SFR-weighted standard deviation of the velocity distribution minimised over $10^4$ viewing angles), which is intended to represent the observed velocity dispersion, versus $\sigma_{Fit}$ (radially averaged velocity dispersion from the best-fitting tilted ring models) for a subset of simulated galaxy snapshots that exhibit disc-like kinematics.
The black dashed line indicates a one-to-one correlation. 
The proxy for velocity dispersion used in this work (\sigoned) generally agrees with the velocity dispersion from detailed disc modelling, an approach used in recent IFS observational work.   
} 
\label{fig:bbsig}
\end{figure}

\subsection{Kinematic properties}
We aim to define a proxy for intrinsic velocity dispersion that we can compare with the statistical trends from recent IFS observations and use to assess the overall kinematic evolution in individual cosmological simulations.
We define a quantity \sigoned\ that is intended to be representative of deriving a flux-weighted mean with some corrections for the beam-smearing effect present in real observations.
We calculate the SFR-weighted standard deviation of the velocity distribution within 0.1 \Rvir\ (typically $\sim10-20$\,kpc) for $10^4$ viewing angles for each galaxy and define \sigoned\ as the minimum value taken over the different viewing angles.
The radius 0.1 \Rvir\ is chosen since it is comparable to the typical field of view of IFS observations.
If we simply calculate a 1-D velocity dispersion without minimising over viewing angles, then this value would be unreasonably large, as it would also include significant large-scale motions, such as rotation (see Table~\ref{tab:propertysim}).
In cases in which clear disc-like kinematics are present, \sigoned\ represents the flux-weighted velocity dispersion measured along a face-on viewing angle.
However, in cases with more complex kinematics, such as late-stage galaxy mergers \citep[e.g.][]{Bellocchi2013}, the physical meaning of \sigoned\ is not as well-defined and only broadly reflects the degree of disturbance of the kinematics.

We caution that \sigoned\ is not directly comparable to the values derived from IFS observations without performing radiation transfer of synthetic spectra, adding instrumental effects, and analyzing the datacubes in the same manner
as real observations. However, we have confirmed that \sigoned\ appears to be an unbiased tracer of the velocity dispersion derived from detailed kinematic modelling:
for a subset of simulated galaxies exhibiting disc-like kinematics, we also derived the velocity dispersion based on modelling of the kinematic maps and datacubes in a manner more akin to observational work.
We used the publicly available code $^{\rm 3D}$Barolo \citep{Di-Teodoro2015,Di-Teodoro2016} to fit tilted ring models to the simulated galaxies in position-position-velocity space.
As shown in Figure~\ref{fig:bbsig}, the radially averaged velocity dispersion derived from the tilted-ring modelling agrees with the cruder \sigoned\ measure to within a factor of two, and there is no obvious systematic bias.

\begin{figure}
\centering
  \includegraphics[width=0.5\textwidth]{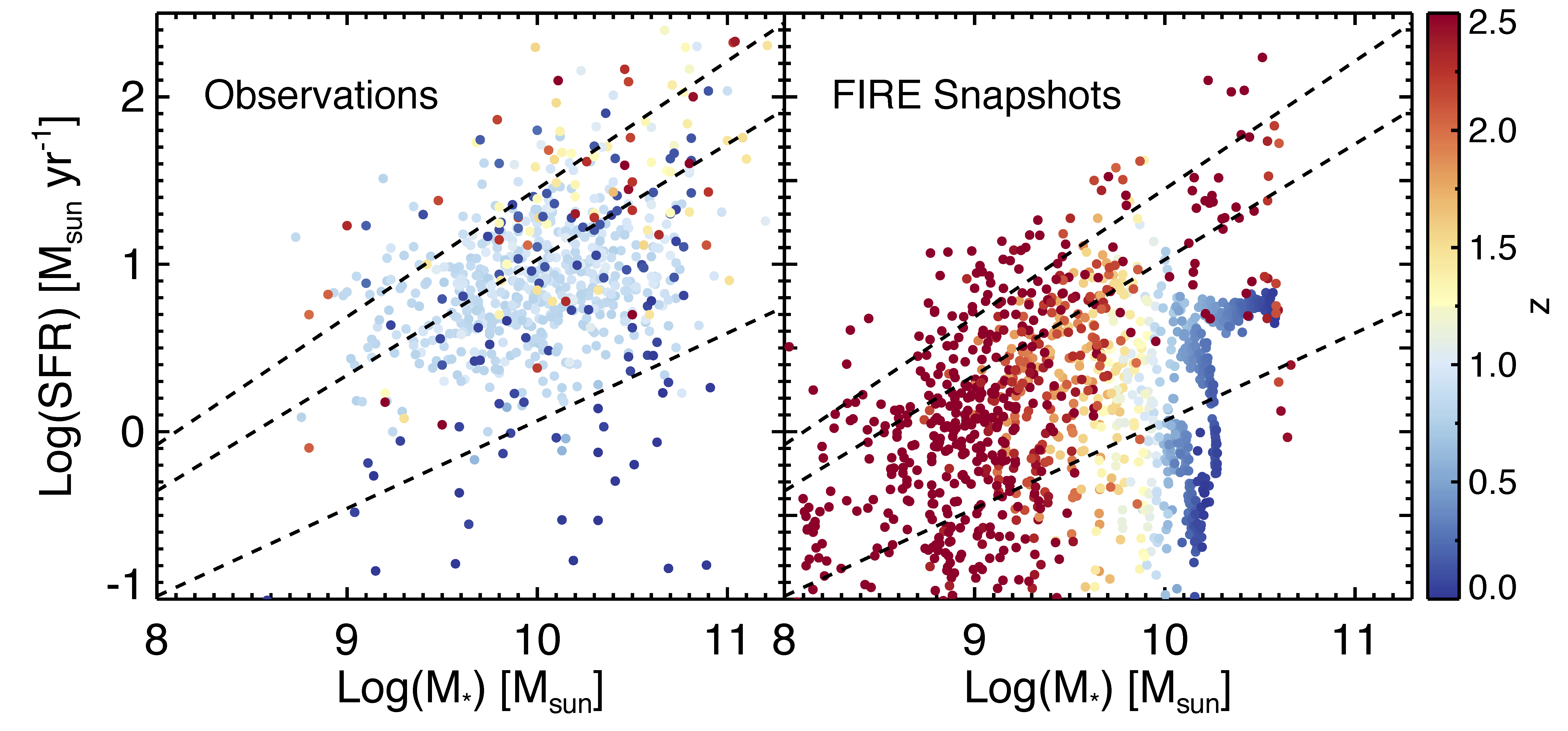} 
\caption{Distribution of the observed and simulated galaxies in the SFR--\Mstar\ plane, colour-coded according to redshift.
The simulated galaxies include all snapshots of the 10 simulations (sampled every few tens of Myr) that fall above the galaxy main sequence.
The SFRs shown in the right panel correspond to the time-averaged SFR calculated based on stars formed within the past 10\,Myr.
The three dashed lines from bottom to top indicate the star-forming galaxy main sequence at $z=0.15$, 1, 2.2 from \citet{Speagle2014}.
The observed and simulated galaxies broadly overlap. In the simulations, the apparent trend of increasing stellar mass
with redshift is an artefact of analysing only three haloes run to $z = 0$, all of which have $z = 0$ halo masses of $\sim 10^{12}$
\Msun.
} 
\label{fig:ms_z}
\end{figure}

\section{A comparison with observations}

\subsection{Samples of simulated and observed galaxies}
We compare the kinematic properties of the simulated galaxies from the FIRE project with observations from the literature. 
From all snapshots of the 10 simulations, we select the subset of central galaxies that are located on or above the star-forming galaxy main sequence \citep[MS;][]{Brinchmann2004,Noeske2007}.
The membership and the distance from the galaxy MS for each galaxy are determined based on the parametric form from \citet{Speagle2014}, with a scatter of 0.2 dex.
The distributions of \Mstar, SFR, and \sigoned\ are summarised in Table~\ref{tab:propertysim}.
We note that large-volume cosmological simulations find that the normalisation of the galaxy MS in simulations is systematically lower than in observations at $z \sim 2$ by up to $\sim0.5$ dex \citep[e.g.][]{Sparre2015,Furlong2015}, and this may also be the case for the FIRE-1 simulations \citep{Sparre2017}, although the small number of haloes analysed in that work prevents a firm conclusion. For this reason, we have checked how sensitive our results are to the definition of the MS; we found that the differences in the median \sigoned\ derived here are typically less than 10 km\,s$^{-1}$ when using different normalisations.

We compile the intrinsic velocity dispersion of star-forming galaxies at $0\lesssim z\lesssim3$ from the literature: GHASP \citep{Epinat2008,Epinat2010}, DYNAMO \citep{Green2014}, the LBA survey \citep{Goncalves2010}, the MUSE/KMOS survey \citep{Swinbank2017}, KROSS \citep{Stott2016,Harrison2017,Johnson2017}, IROCKS \citep{Mieda2016}, MASSIV \citep{Epinat2012,Queyrel2012}, WiggleZ \citep{Wisnioski2011}, KLASS \citep{Mason2017}, SHiZELS \citep{Molina2016}, CASSOWARY \citep{Leethochawalit2016}, SINS disc-like galaxies \citep{Cresci2009}, lensed galaxies \citep{Livermore2015}, the \citet{Law2009} OSIRIS survey, KDS \citep{Turner2017}, and VVDS \citep{Lemoine-Busserolle2010a}.
The kinematic properties derived from these surveys are based on nebular emission lines (e.g. \Ha, \OIII, and \OII) and thus trace ionised gas.
Because we compute \sigoned\ of the gas particles in the simulated galaxies using an SFR weighting, this measure should be roughly analogous to that derived from nebular emission lines in observational surveys \citep{Hung2016}.
Here, we only include measurements from IFS surveys (either AO-assisted or seeing-limited observations) that are based on spatially resolved velocity dispersion maps and are limited to including those studies that made their measurements available.
We do not include dispersion measurements based on integrated spectra since they likely include bulk motions, such as rotation \citep[e.g.][]{Forster-Schreiber2009}.
The median $z$, \Mstar\ (when available), SFR (when available), and \sigoned\ values of the star-forming galaxies in each survey are summarised in Table~\ref{tab:propertyobs}.

Figure~\ref{fig:ms_z} illustrates the distribution of the simulated and observed galaxies in the SFR--\Mstar\ plane, colour-coded according to redshift.
For the three halos run to $z = 0$, a clear trend of increasing \Mstar\ with decreasing redshift is seen for the simulated galaxies as a direct result of continuous mass assembly over time.
However, due to the small number of simulated galaxies analysed here, the dynamic range in \Mstar\ at a given redshift is significantly less than that spanned by the observed galaxies.

\begin{figure}
\centering
  \includegraphics[width=0.5\textwidth]{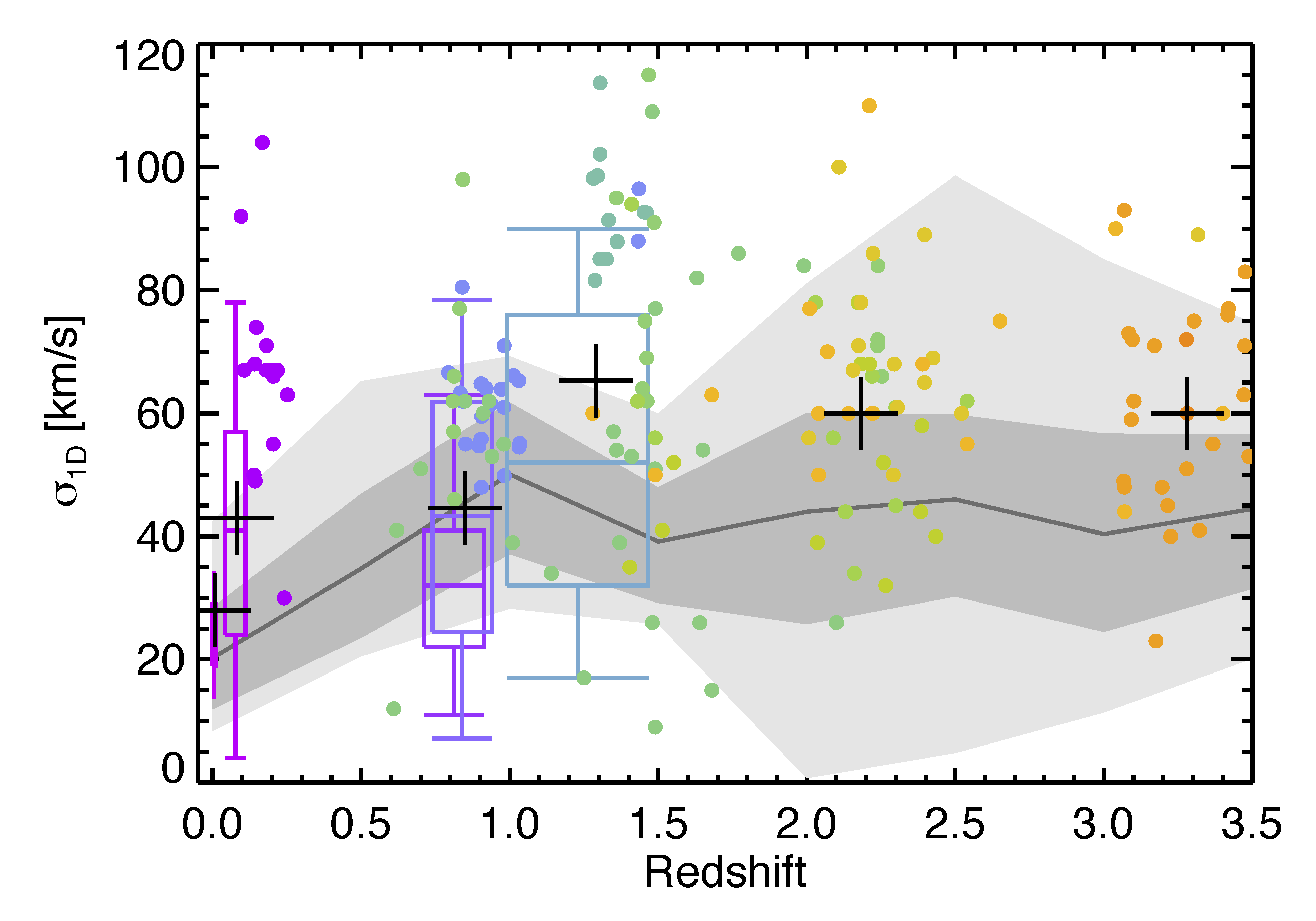} 
\caption{\sigoned\ as a function of redshift, similar to the observational results compiled recently by \citet{Wisnioski2015} and \citet{Turner2017}.
The dark grey solid line shows the median value for the simulated galaxies that lie on or above the star-forming galaxy MS within redshift bins of width $\Delta z=$0.5 (as listed in Table~\ref{tab:propertysim}),
and the lighter (lightest) grey coloured area encloses 68 per cent (95 per cent) of the data.
The simulated galaxies include all snapshots of the 10 simulations sampled every few tens of Myr.
The coloured dots and the distribution bars summarise the observations described in Section 3.1; each colour represents an IFS survey.
In each distribution bar, the middle bar corresponds to the median \sigoned, and the boxes and vertical bars encompass 68 and 95 per cent of the data, respectively.
The width of each distribution bar is proportional to the redshift coverage of the survey.
The black crosses represent the median \sigoned\ of the observed galaxies that fall on or above the galaxy MS in six redshift bins: $z<0.1$, $0.1<z<0.3$, $0.7<z<1.0$, $1.0<z<1.5$, $2.0<z<2.5$, and $3.0<z<3.5$.
The simulated and observed galaxies exhibit the same qualitative trend, i.e. the median \sigoned\ increases from $z = 0$ to $z \sim 1$ and then remains constant. 
} 
\label{fig:sigz}
\end{figure}

\begin{figure}
\centering
  \includegraphics[width=0.5\textwidth]{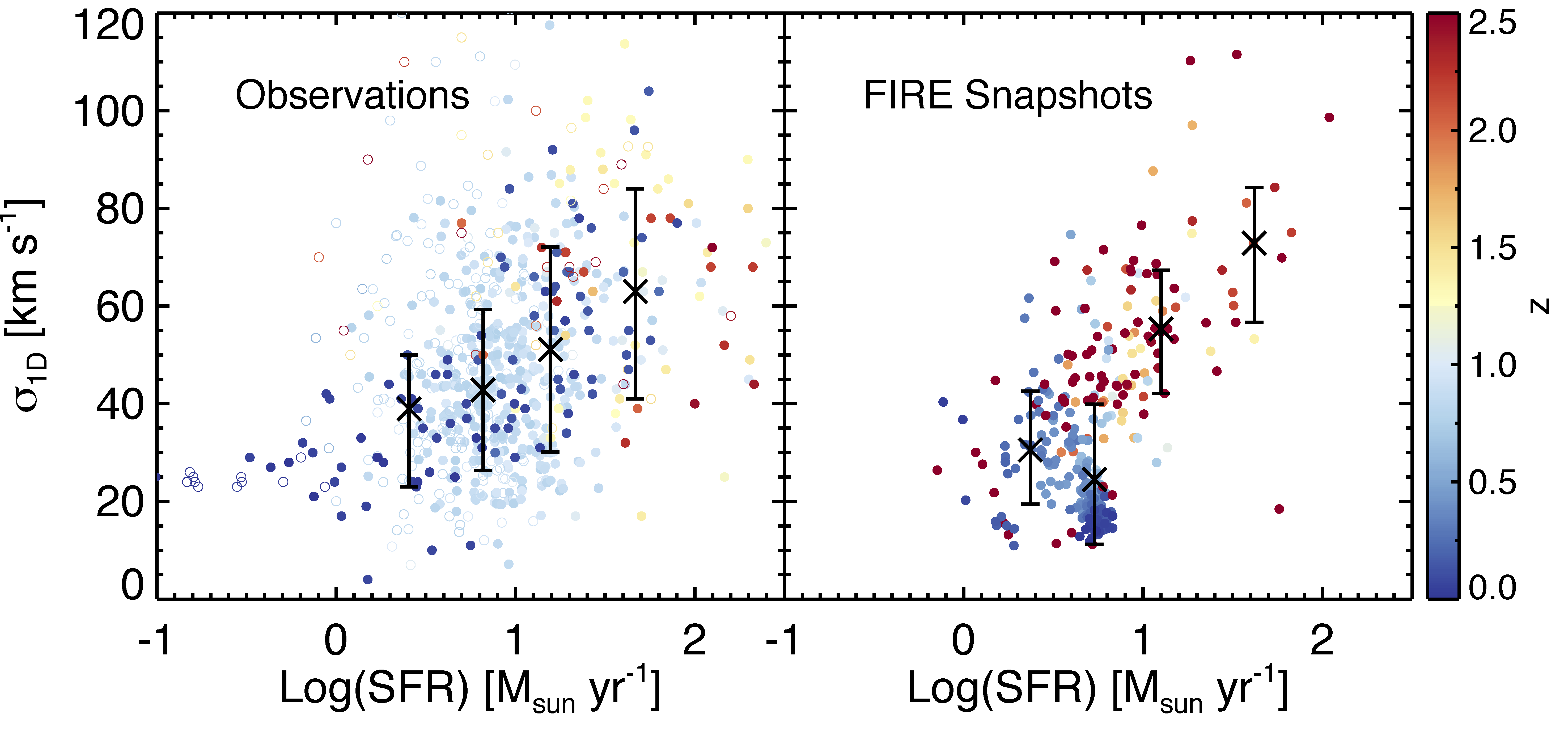} 
\caption{Distributions of the observed and simulated galaxies in the \sigoned--SFR plane, coloured-coded according to redshift.
The open circles in the observations panel indicate galaxies that fall below the galaxy MS.
The black x's and error bars overlaid on the data points represent the median \sigoned\ and 68 per cent distribution of the galaxies that lie on or above the MS in four $\log ({\rm SFR})$ bins.
The simulations and observations generally both exhibit positive correlations between \sigoned\ and SFR.
} 
\label{fig:sigsfr}
\end{figure}

\begin{figure}
\centering
  \includegraphics[width=0.5\textwidth]{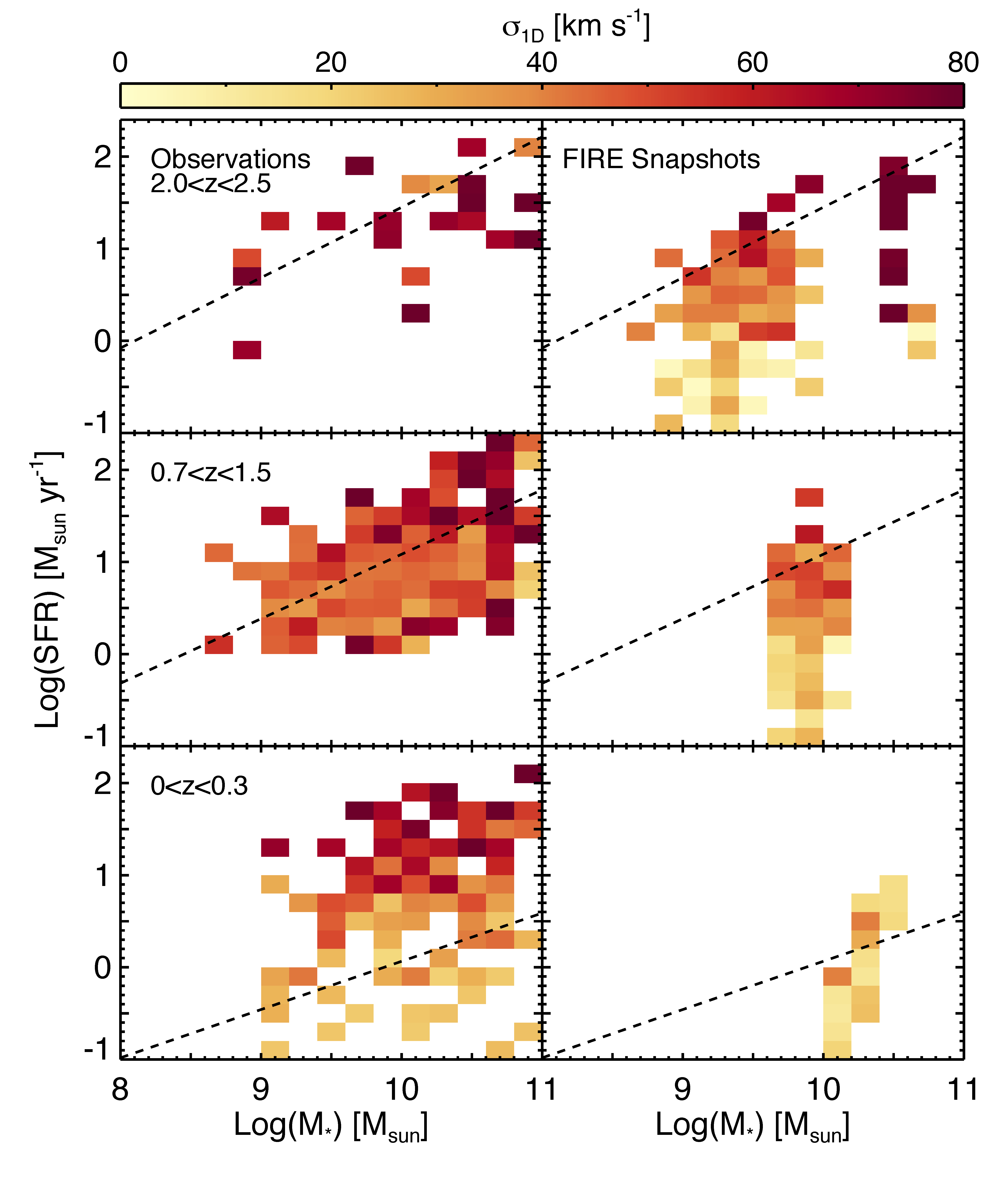} 
\caption{Distributions of the observed and simulated galaxies in the SFR--\Mstar\ plane in three redshift bins ($2<z<2.5$, $0.7<z<1.5$, and $0<z<0.3$), colour-coded according to \sigoned.
Each small \Mstar\ and SFR bin ($\Delta \log$(\Mstar)$=0.2$ and $\Delta \log$(SFR)$=0.2$) is colour-coded based on the median \sigoned\ values for all galaxies within the bin.
The three dashed lines in the top, middle, and bottom panels indicate the star-forming galaxy MS at $z=2.2$, 1, 0.15. In some panels, \sigoned\ tends to increase with SFR, whereas in others,
\sigoned\ seems to be correlated in \Mstar, and in other panels, no clear trend is observed.
It is worth noting that especially at high redshift, the simulated galaxies tend to include galaxies with SFRs well below the MS, unlike the observed galaxy samples, likely because of
observational selection effects.
} 
\label{fig:mssig}
\end{figure}
 
\subsection{Observational diagnostics:~\sigoned\ as a function of $z$ and SFR}
In this section, we examine whether the FIRE simulations exhibit the general trends in \sigoned, $z$ and SFR present in the observations.
Figure~\ref{fig:sigz} shows \sigoned\ as a function of redshift, as summarised in Tables~\ref{tab:propertysim} and~\ref{tab:propertyobs}.
The median \sigoned\ values of the observed galaxies are generally $\sim10-20$ km\,s$^{-1}$ greater than those of the simulated galaxies.
This systematic offset may be a result of several factors: the fact that \sigoned\ cannot perfectly replicate the measured intrinsic velocity dispersion, the imperfect (or lack of) beam-smearing corrections in the IFS measurements \citep[e.g.][]{Stott2016,Pineda2017}, and/or differences in physical properties such as SFR and \Mstar\ (Figure~\ref{fig:ms_z}).
Regardless, the observed and simulated samples follow a similar trend with redshift, in that the median \sigoned\ increases from $z\sim0$ to $z\sim1$ and then remains approximately constant at $z\gtrsim1$.
The increasing trend from $z\sim0$ to $z\sim1$ is consistent with the results of \citet{Kassin2014}, who analysed the evolution of the velocity dispersion traced by cold and warm gas in a suite of cosmological zoom simulations. 
We note that the observational data compiled here do not exhibit a monotonically increasing trend with redshift, as concluded by \citet{Wisnioski2015}, which is a result of the $\sim20-30$ km\,s$^{-1}$ lower velocity dispersion in the $z\sim1$ KMOS$^{\rm 3D}$ sample compared with other galaxy surveys at the same epoch that include galaxies with comparable SFR and \Mstar\ values (e.g. KROSS, IROCKS, MASSIV, and WiggleZ).
Whether \sigoned\ continues to rise beyond $z>1$ also depends on which surveys or subset of the data are included.

Figure~\ref{fig:sigsfr} shows the distribution of \sigoned\ and SFR for the observed and simulated galaxies.
For the observed galaxies, a positive correlation is seen out to $z\lesssim1$, consistent with the results of \citet{Green2014}.
No obvious offset is seen in the \sigoned-SFR relation within the redshift range probed.
However, it is less clear whether the same positive correlation extends to $z\gtrsim2$ galaxies owing to the small number of galaxies.
The star-forming galaxies (galaxies that lie on or above the MS) from the FIRE simulations also exhibit a positive correlation between \sigoned\ and SFR, consistent with the observed galaxies.
There is also no obvious offset in the \sigoned-SFR relations at different redshifts except for a group of low-redshift snapshots from the m12i simulation that lie significantly below the relations.

Although the FIRE simulations reproduce the observed trends in the \sigoned--$z$ and \sigoned--SFR planes, understanding what physical processes are responsible for these trends remains challenging.
First, the interpretation that  velocity dispersion increases with redshift is tangled with the positive correlation between \sigoned\ and SFR, given that galaxies targeted by IFS surveys at $z\gtrsim1$ tend to have higher
SFRs compared with galaxies targeted by lower-redshift IFS surveys.
Second, the \sigoned\ values of the observed galaxies exhibit large scatter even after taking into account the differences in SFR and removing galaxies that fall below the MS, suggesting that star formation is
not the only relevant driver of velocity dispersion (although this interpretation can be complicated if the SFR and \sigoned\ vary on short timescales and are not perfectly synchronised; this is explored in detail
below).
The scatter in the \sigoned-SFR relation likely cannot be attributed to different techniques or definitions used in each survey given that there is a similarly large scatter in the simulated galaxies.
Several previous works have examined the relationship between velocity dispersion and SFR surface density ($\Sigma_{\rm SFR}$) because such a relation may naturally arise from the scaling between $\Sigma_{\rm SFR}$ and gas surface density ($\Sigma_{\rm gas}$) in marginally stable discs \citep{Swinbank2012a}.
However, the \sigoned-$\Sigma_{\rm SFR}$\footnote{Here, we estimate $\Sigma_{\rm SFR}$ of the FIRE galaxies as SFR/$A$, where $A$ is the area calculated based on an effective radius that encloses half of the total SFR.} relation of the FIRE galaxies exhibits a comparably large scatter as seen in the \sigoned-SFR relation.

\subsection{Distribution of \sigoned\ in the SFR--\Mstar\ plane}
In Figure~\ref{fig:mssig}, we plot the distribution of \sigoned\ in the SFR--\Mstar\ plane for both observed and simulated galaxies in three redshift bins.
These figures demonstrate that how \sigoned\ depends on galaxy properties varies with redshift, thus complicating interpretation of these results.
Whereas \sigoned\ tends to increase with SFR in the lowest-redshift panel for the observed galaxies and the two higher-redshift panels for the simulated galaxies, this trend with SFR is less evident for the observed galaxies in the two higher-redshift bins.
Another possible trend of increasing \sigoned\ with \Mstar\ has been reported in \citet{Stott2016}, but this trend no longer exists (or exists only at the high-mass end for the full sample) in the updated analysis in \citet{Johnson2017}.
This is evident in the $0.7<z<1.5$ panel for the observed galaxies (mostly based on KROSS galaxies) and the $2.0<z<2.5$ panel for the simulated galaxies.
This potential trend with \Mstar\ suggests that variations in \Mstar\ within samples may contribute to the large scatter in the \sigoned-SFR relation of a given sample.
However, no clear trend with \Mstar\ is seen in any other panel.

Finally, whereas most observed galaxies with SFR $\gtrsim10$ \Msun\,yr$^{-1}$ at all redshifts have \sigoned\ $\gtrsim50$ km\,s$^{-1}$, there is a clear discrepancy between the $0<z<0.3$ and $0.7<z<1.5$ bins for the lower-SFR galaxies.
For SFR $\lesssim10$ \Msun\,yr$^{-1}$, galaxies at $z\sim1$ have higher \sigoned\ than their lower-redshift counterparts at a given SFR and \Mstar.
Although the overlap in SFR and \Mstar\ is limited for the simulated galaxies in different redshift bins, a slight increase in \sigoned\ with redshift is seen in the regions of overlap between the $z\sim2$ and $z\sim1$ panels and between the $z\sim1$ and $z\sim0$ panels.
These results suggest that for at least some (lower-SFR) galaxies, the intrinsic velocity dispersion varies with redshift even for galaxies with similar \Mstar\ and SFR.

\begin{figure*}
\centering
  \includegraphics[width=\textwidth]{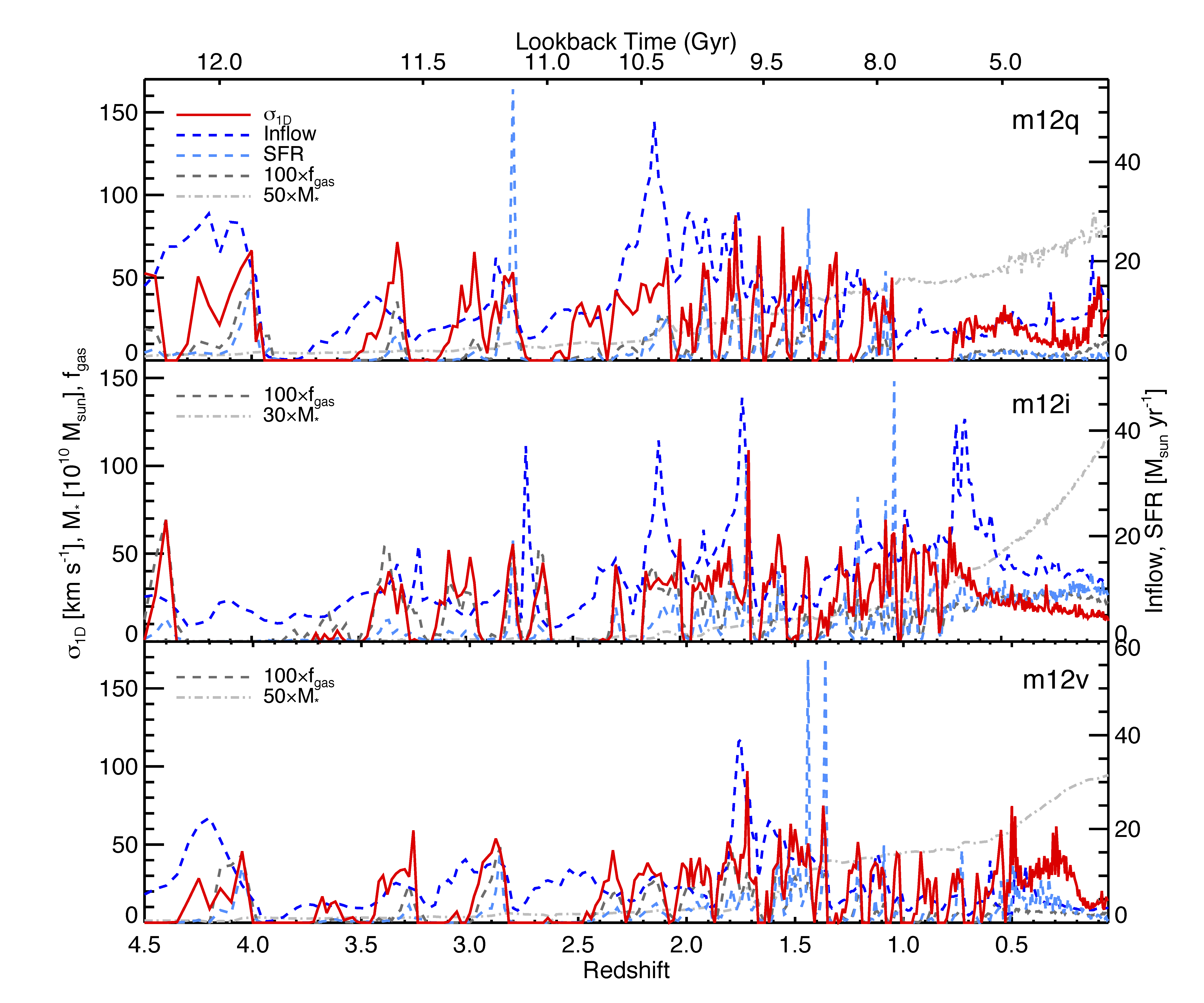} 
\caption{Time evolution of \sigoned\ (red solid lines), instantaneous SFR (light blue dashed lines), gas inflow rate (\Minflow, blue dashed lines), $f_{\rm gas}$ (dark grey dashed lines), and \Mstar\ (grey dash-dot lines) for the three m-series simulations.
$f_{\rm gas}$ is defined as the ratio of the `dense' gas mass ($n\geq\,1\,$cm$^{-3}$) to the sum of the gas and stellar masses within 0.2 \Rvir.
\Minflow\ is defined as the instantaneous mass flux through a thin spherical shell with outer radius 0.3 \Rvir\ and inner radius 0.2 \Rvir.
The values of $f_{\rm gas}$ and \Mstar\ are arbitrarily scaled (as indicated in the legend) for clarity, and the values of \Minflow\ and SFR are shown on the right axis. \sigoned, SFR, and \Minflow\ vary drastically
on short ($< 100$ Myr) timescales owing to stellar feedback-driven burstiness and stochasticity in gas inflow, which is partially due to galactic fountains. It is clear that peaks in \sigoned, \Minflow, and SFR
tend to temporally coincide but are not exactly simultaneous. The times at which \sigoned\ approaches zero correspond to times when few star-forming gas particles are present in the simulations.
} 
\label{fig:m12qq_siglos}
\end{figure*}

\begin{figure}
\centering
  \includegraphics[width=0.49\textwidth]{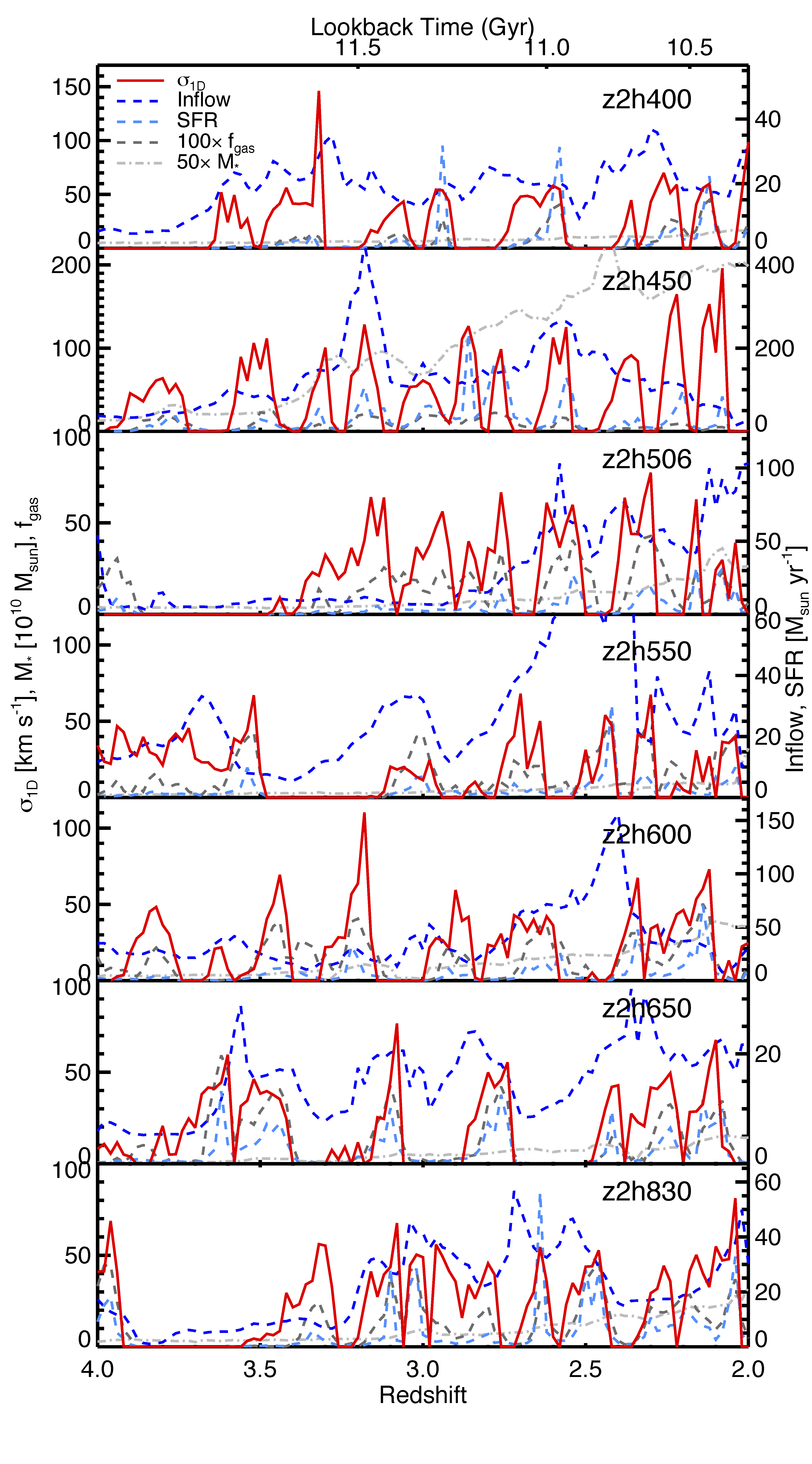} 
\caption{Time evolution of \sigoned\ (red solid lines), instantaneous SFR (light blue dashed lines), \Minflow\ (blue dashed lines), $f_{\rm gas}$ (dark grey dotted lines), and \Mstar\ (grey dash-dot lines) of the seven z2h simulations.
The values of $f_{\rm gas}$ and \Mstar\ are arbitrary scaled (as indicated in the legend) for clarity, and the values of \Minflow\ and SFR are shown on the right axis. These higher-mass galaxies simulated
only to $z = 2$ exhibit the same qualitative behaviour as the lower-mass galaxies shown in Figure~\ref{fig:m12qq_siglos}.
} 
\label{fig:z2h_siglos}
\end{figure}

\section{Physical drivers of kinematic evolution} \label{sec:drivers}
To gain further insight into the physical drivers of galaxy dynamics, we analyse the time evolution of various physical properties of the individual central galaxies of each halo.
Figures~\ref{fig:m12qq_siglos} and~\ref{fig:z2h_siglos} show the time evolution of \sigoned, SFR, gas inflow rate (\Minflow), gas fraction ($f_{\rm gas}$), and \Mstar\ of the three m-series and seven z2h simulations,
respectively.
Several recent studies have investigated the star formation, gas accretion, and mass assembly histories of these FIRE simulations in detail \citep[e.g.][]{Hopkins2014,Muratov2015,Angles-Alcazar2017,Sparre2017}.
In these simulations, the gas inflow rate varies on timescales of a few 100 Myr.
Although several major mergers lead to dramatic increases in \Minflow, minor mergers and smooth gas accretion, partially due to galactic fountains, predominantly govern the amount of material for subsequent star formation.
The SFRs of these galaxies are highly dynamic, and individual galaxies can evolve through quiescent and star-forming phases within a few 100 Myr \citep{Sparre2017}.
The gas fractions vary on similar timescales as SFR and \Minflow\ whereas \Mstar\ tend to vary on longer timescales.
Recall that the gas fraction is computed as the ratio of the dense ($n\geq\,1\,$cm$^{-3}$) gas mass to the total stellar and dense gas mass within 0.2 $R_{\rm vir}$. 
We note that the behavior of gas fractions can change significantly depending on the density threshold that we employ. 
A lower or even no threshold would likely include a significant amount of gas outside of the ISM/galaxy disc (i.e. the circumgalactic medium, including gas in galactic fountains; \citealt{Muratov2015,Angles-Alcazar2017}); consequently, the gas mass in the ISM may exhibit short or long timescale variations that are not reflected in other gas fraction measures. 
The gas fraction in the ISM/disc is the relevant quantity for self-regulated star formation models \citep[e.g.][]{Thompson2005,Ostriker2011, Faucher-Giguere2013,Hayward2017}. 
A more detailed comparison with the predictions of such models should use a proxy that better represents the disc fraction, but we defer such a comparison
to future work and simply advise that the time evolution of the gas fraction plotted here should be interpreted with the above caveat in mind.

Among the properties shown in Figures~\ref{fig:m12qq_siglos} and \ref{fig:z2h_siglos}, the time evolutions of SFR and \Minflow\ most closely resemble the variations in \sigoned; all three
properties vary significantly on timescales of $<100$ Myr.
By definition, \sigoned\ approaches zero at low SFR (SFR $\lesssim$ 1 \Msun\,yr$^{-1}$), since only a few SPH gas particles with non-zero SFRs are available to trace the galaxy dynamics.
In general, such time periods also correspond to when galaxies have lower gas inflow rates.
Increases in \sigoned\ occur near enhancements in \Minflow\ and SFR, although not exactly simultaneously.
The approximate temporal correspondence of variations in \sigoned\ and SFR is reflected in the positive SFR--\sigoned\ correlation, and the fact that peaks do not exactly coincide in time may be responsible for
the large scatter in this relation (Figure~\ref{fig:sigsfr}).
The similarities between the evolutions of SFR, \sigoned, and \Minflow\ suggest that variations in velocity dispersion, star formation (and thus prompt stellar feedback) and gas inflow are all related.

\begin{table}
\centering
 \caption{Time delay measurements for the simulations}
 \label{tab:tdelay}
 \resizebox{0.45\textwidth}{!}{%
 \begin{tabular}{@{}lcccc}
 \hline
 \hline
Simulation & Range & $\tau$(\Minflow-SFR)$^{a}$ & $\tau$(\Minflow-\sigoned) & $\tau$(\sigoned-SFR) \\
          & (Gyr)  &   (Myr)   & (Myr) & (Myr) \\
 \hline
m12q & 0$-$12 & -69$^{b}$ & 0 & 0   \\
m12i & 0$-$12 & -100 & 0 & -34  \\
m12v  & 0$-$12 & -34 & -35 & 0   \\
z2h400 & 10.5$-$12 & 331$^{*}$ & 0$^{*}$ & -17 \\
z2h450 & 10.5$-$12 & -331$^{*}$ & 488$^{*}$ & 0 \\
z2h506 & 10.5$-$12 & -52 & 418 & 0$^{*}$ \\
z2h550 & 10.5$-$12 & -87 & -87 & 0 \\
z2h600 & 10.5$-$12 & -348 & -348 & 0 \\
z2h650 & 10.5$-$12 & -296$^{*}$ & -35 & 0 \\
z2h830 & 10.5$-$12 & -70 & 104$^{*}$ & 0 \\
 \hline
 \multicolumn{5}{l}{$^a$Time delay measurements based on cross-correlation analyses }\\
 \multicolumn{5}{l}{~~between gas inflow rate and instantaneous SFR ($\tau$(\Minflow-SFR)), }\\
 \multicolumn{5}{l}{~~gas inflow rate and \sigoned\ ($\tau$(\Minflow-\sigoned)), }\\
 \multicolumn{5}{l}{~~and \sigoned\ and instantaneous SFR ($\tau$(\sigoned-SFR)).} \\
\multicolumn{5}{l}{$^b$Peak value in the $\pm$500 Myr window.}\\
\multicolumn{5}{l}{$^*$Multiple comparable peaks within the $\pm$500 Myr window.}\\
 \end{tabular}}
\end{table}


To examine whether the similarities in the temporal evolutions of SFR, \sigoned, and \Minflow\ are indicative of any causal effects, we measure the time delays between pairs of these three quantities
by cross-correlating their time series.
Because these zoom-in simulations have unequal time intervals, we first bin and interpolate these time series with a uniform sampling rate determined based on the mean time interval of the entire time series ($\sim34$ Myr for the m-series and $\sim17$ Myr for the z2h simulations).
For each simulation suite, we calculate temporal cross-correlation functions of pairs of time series [$(x,y) = ($\Minflow, SFR), (\Minflow,\sigoned), and (\sigoned, SFR)]
that are normalised to have means of 0 and variances of 1, CCF$(\tau_k)=\frac{1}{N}\sum_{i=1}^{N-k}x_i y_{i+k}$, where $i$ indicates the time bin, $N$ is the total number
of time bins, $\tau_k$ is the lag corresponding to the $k$ time bin, and $k$ is varied to probe lags from $-500$ to 500 Myr.
Table~\ref{tab:tdelay} summarises the time delay measurements ($\tau$(\Minflow-SFR), $\tau$(\Minflow-\sigoned), $\tau$(\sigoned-SFR)) by taking the peak value of CCF$(\tau_k)$ within a $\pm500$-Myr.
Values in Table~\ref{tab:tdelay} marked with a ``$*$'' are not robust because there are multiple comparable peaks within a $\pm500$ Myr window.
 
We note that the absolute values of these time delay measurements can change by a factor of a few when employing different time sampling rates.
However, the signs of these time delay measurements are relatively stable.
In general, variations in \Minflow\ occur prior to variations in SFR, suggesting that inflows through 0.2 \Rvir, due to cosmological gas inflow or galactic fountains, lead to subsequent enhanced star formation activity.
As expected, the relation between \Minflow\ and SFR becomes weaker if we define the inflow rate as the mass infalling through a shell with a radius of 1 \Rvir.
Negative signs for $\tau$(\Minflow-\sigoned) or $\tau$(\sigoned-SFR)  are obtained for a few simulated galaxies, suggesting that enhancements in velocity dispersion may occur after enhancements in the gas inflow rate at 0.2 \Rvir\ but before the SFR is enhanced.
However, these trends are less robust because positive or zero time delays are also obtained for some simulated galaxies.
Given that the absolute values of $\tau$(\Minflow-\sigoned) and $\tau$(\sigoned-SFR) are typically less than the resolution of the time series, we can only conclude that variations in \sigoned\ roughly temporally coincide with variations in \Minflow\ and SFR.
On a related note, \citet{El-Badry2016,El-Badry2017} investigated the stellar kinematics of low-mass galaxies from the FIRE project and found that the stellar velocity dispersion is strongly correlated with SFR, with a $\sim50$\,Myr time-delay.
They attribute this correlation to both the SFR and stellar velocity dispersion being affected by stellar feedback and the consequent gas outflows.

\section{Discussion}
\subsection{Comparison with previous work}
Several previous work have developed models that aim to examine the roles of gas accretion, gravitational instabilities, stellar feedback, and other physical processes in driving the enhanced velocity dispersions of high$-z$
galaxies. Comparisons between models and observations can provide some insights into the physical drivers since these models predict different relationships between $\sigma$ and other physical parameters.
For example, simply requiring a marginally stable disc with Toomre $Q$ parameter $\approx1$, we expect $\sigma \propto f_{\rm gas}$ \citep[e.g.][]{Thompson2005,Faucher-Giguere2013},
with no (additional) dependence on SFR (this \emph{does not}
preclude a correlation between \sigoned\ and SFR because all else being equal, owing to the Kennicutt-Schmidt relation \citep{Kennicutt1998,Schmidt1959}, SFR and $f_{\rm gas}$ should be correlated).
A recent work by \citet{Krumholz2016} argues that in the steady-state configuration of the gravity-driven, turbulent disc model from \citet{Krumholz2010}, SFR\,$\propto f^{2}_{\rm gas} \sigma$. 
This work also argues that models in which stellar feedback drives turbulence and ``self-regulates'' galaxy-scale star formation \citep[e.g.][]{Thompson2005,Ostriker2011,Shetty2012,Faucher-Giguere2013,Hayward2017}
predict a relation of SFR\,$\propto\sigma^2$ with no dependence on $f_{\rm gas}$ \citep[eq. 8 of][]{Krumholz2016}, but c.f. \citealt{Ostriker2011} and \citet{Hayward2017} for alternative interpretations.

In principle, the FIRE simulations directly include the relevant processes (e.g. gravitational instability, gas inflow, and stellar feedback-driven turbulence), and the similarities between the
\sigoned, SFR, and \Minflow\ time series demonstrated in Section \ref{sec:drivers} may indicate that one or both of these proposed mechanisms are at work.
A positive correlation between SFR and \sigoned, consistent with the expectations of both gravity- and feedback-driven turbulence models, is seen in both observations and FIRE simulations (Figure~\ref{fig:sigsfr}).
The time variation of $f_{\rm gas}$ resembles that of the SFR (Figures~\ref{fig:m12qq_siglos} and~\ref{fig:z2h_siglos}), and thus we may expect some dependence between $f_{\rm gas}$ and \sigoned\ as well.
However, we caution against a direction comparison of this dependence with analytic models because in many snapshots (especially at high redshift), the star-forming gas does not exhibit disc-like kinematics, which is inconsistent with the assumptions of many analytic models. Moreover, as discussed above, the definition of $f_{\rm gas}$ in this work is not completely analogous to that in the aforementioned models.
Finally, the significant time variability in e.g. SFR implies that such quasi-steady-state models may only hold in a time-averaged sense;
see \citet{Hayward2017}, \citet{Torrey2017}, and \citet{Faucher-Giguere2018} for further details.

Some analytic models have argued that accretion energy itself is unable to drive turbulence for longer than a characteristic accretion time \citep[on the order of a few hundred Myr; e.g.][]{Elmegreen2010}, but others
have argued that the following inflow within the disc can drive turbulence for much longer \citep{Genel2012}.   
Moreover, an important characteristic of the simulated galaxies is that $\sigma$ can vary significantly on timescales of $<100$ Myr, including ``gaps'' of comparable time for which $\sigma$ of the
star-forming gas is not traceable owing to a lack of star-forming gas particles.\footnote{This can occur if no gas particles satisfy the criteria for star formation used in the FIRE simulations;
see Section \ref{sec:fire_methods}.}
This is in part due to the onset of stellar feedback processes such as radiation pressure, stellar winds, and supernovae, which can disrupt dense clumps and expel a significant fraction of
the ISM \citep{Muratov2015,Hayward2017}, causing star formation to temporarily cease and thus making it impossible to trace the gas kinematics via nebular emission lines.
In this situation, mechanisms that can drive enhancements in the velocity dispersion of star-forming gas do not necessarily need to maintain enhanced velocity dispersions over timescales
longer than a few hundred Myr. Thus, we cannot rule out the importance of accretion in driving $\sigma$ simply because the impact of accretion may be relatively short-lived.
Overall, our results demonstrate the difficulty of inferring the physical driver(s) of velocity dispersion from integrated measurements, such as the global SFR, mainly because
enhancements in \sigoned, \Minflow, and SFR are not perfectly temporally coincident, which results in a significant \emph{physical} scatter in the relationships amongst these quantities.

\begin{figure}
\centering
  \includegraphics[width=0.5\textwidth]{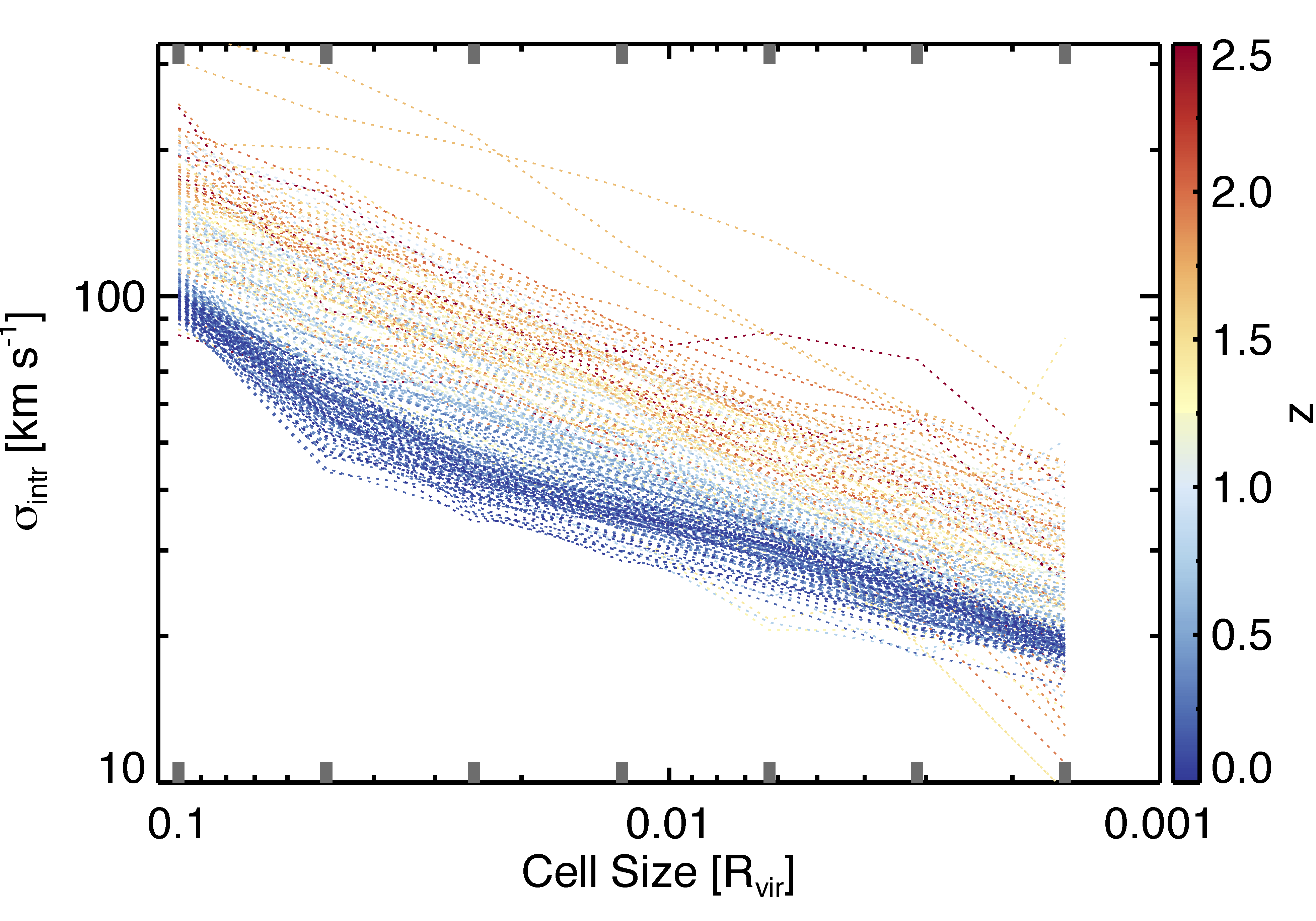} 
\caption{Local velocity dispersion (\sigintr) versus cell size (from large to small) for each snapshot of the m12i simulation, colour-coded according to redshift.
The redshift dependence is complicated by the fact that \Rvir\ itself is a function of redshift. 
\sigintr\ is calculated based on all gas particles with temperature $<10^4$ K and within a distance of 0.1 \Rvir\ from the halo centre.
The thick grey bars indicate the discrete cell sizes we use to calculate \sigintr.
Within the range of cell sizes probed, \sigintr\ decreases with decreasing cell size; the relationship is roughly consistent with the observed
linewidth-size relation \citep{Larson1981}.
} 
\label{fig:m12i_turball}
\end{figure}

\begin{figure}
\centering
  \includegraphics[width=0.5\textwidth]{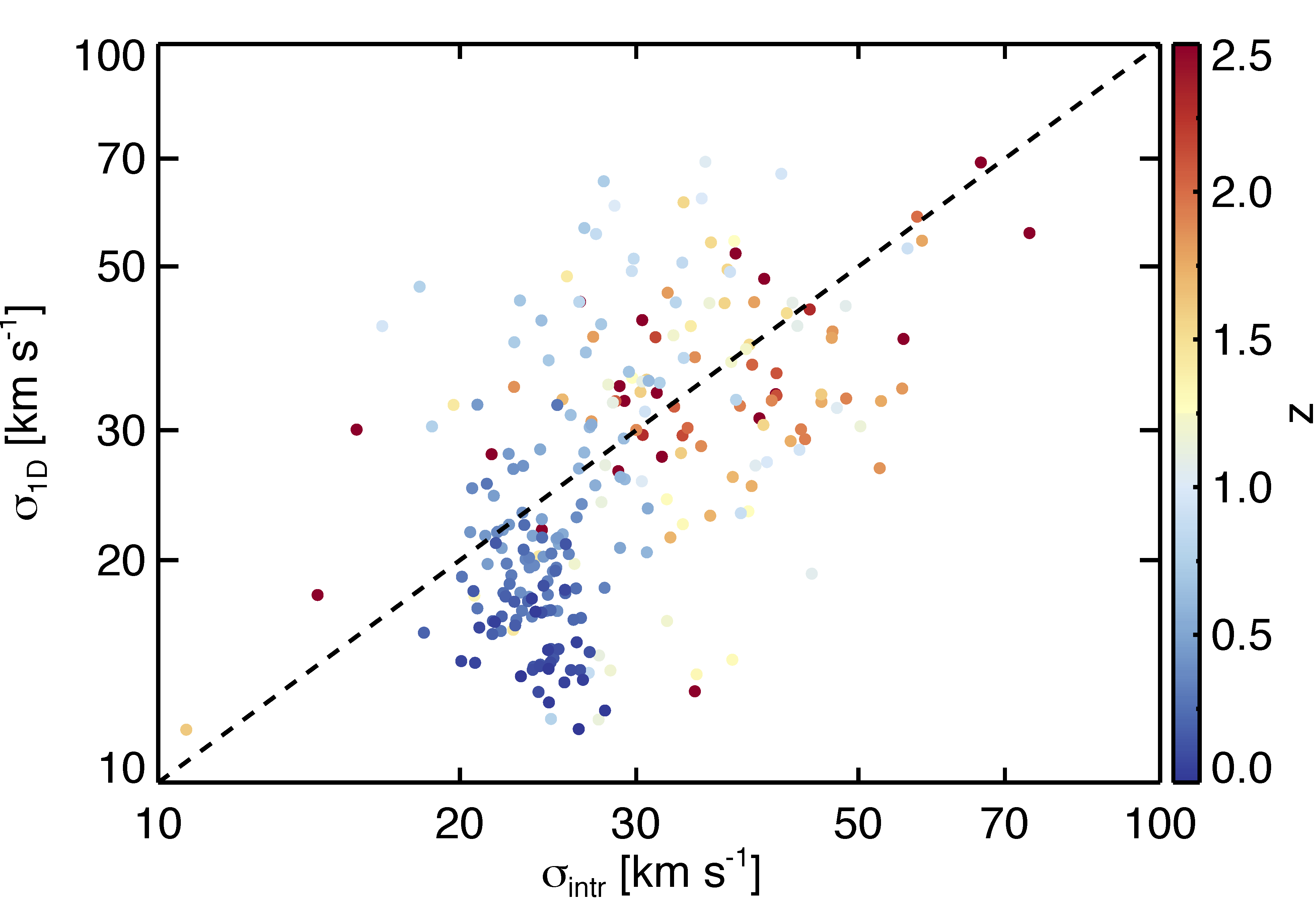} 
\caption{\sigoned\ (SFR-weighted standard deviation of the velocity distribution minimised over $10^4$ viewing angles), which is intended to represent the observed velocity dispersion, versus \sigintr\ (mean velocity fluctuation over all cells with cell size of $\approx0.003$ \Rvir) for the m12i simulation, colour-coded according to redshift.
The black dashed line indicates a one-to-one correlation. The ``observed'' and ``true'' velocity distribution tend to agree, although there is significant scatter in the correlation.
There is a systematic offset at lower redshift, possibly due to \sigintr\ including some differential rotation on kilosparsec scales, whereas \sigoned\ minimises this.
} 
\label{fig:m12i_turbsig}
\end{figure}

\subsection{Physical interpretation of \sigoned}
Whereas the velocity dispersion has been the primary parameter of interest, its physical interpretation can be ill-defined in the absence of a clear, disc-like kinematic structure.
Throughout this work, we have used \sigoned\ as a proxy of velocity dispersion measured in observational work (e.g. through kinematic modelling).
Here, we test how well \sigoned\ truly represents the local random motions of the gas.
We include all gas particle with temperature $<10^4$ K and within a distance of 0.1 \Rvir\ from the halo centre, but we note that our results do not change significantly when we include gas particles with temperature up to $\sim 2 \times 10^4$ K. 
To estimate the amplitude of local random motions, we calculate the mean velocity field over fixed-sized cells (seven different cell sizes are used, with total cell numbers of 1$^{3}$, 2$^{3}$, 4$^{3}$, 8$^{3}$, 16$^{3}$, 32$^{3}$, and 64$^{3}$ sampling the volume within 0.1 \Rvir).
We focus our discussion on only the variation of \sigintr\ across different scales of individual galaxy snapshots and do not address the relative ``trends'' with redshift, as they are complicated by the fact that \Rvir\ itself is a function of redshift.
We calculate the velocity dispersion within each cell with more than 10 gas particles by subtracting the local mean velocity in each cell.
Then, we define a measure of the local random velocity (\sigintr) as the mean velocity dispersion taken over all cells.
Similar approaches have been used to estimate turbulent gas motions in SPH galaxy cluster simulations \citep[e.g.][]{Dolag2005} and galaxy-scale simulations \citep[e.g.][but they use adaptive cell sizes and work in spherical coordinates for galaxies without a well-defined disc]{Su2016}.
Unlike \citet{Su2016}, we do not exclude high-velocity particles that may be outflows driven by stellar feedback, since estimating the pure turbulent energy is not our primary goal.

Figure~\ref{fig:m12i_turball} shows \sigintr\ as a function of cell size for the m12i simulation, colour-coded according to redshift.
A clear decreasing trend between \sigintr\ and decreasing cell size is seen for all snapshots, and it roughly follows a power-law relation consistent with the observed
linewidth-size relation \citep{Larson1981}.
In principle, we could construct e.g. density PDFs and power spectra from the simulations to more thoroughly analyse the turbulence, but such a detailed comparison is beyond the scope of this work.
Instead, we simply compare \sigoned\ with \sigintr\ calculated with a cell size of $\sim0.003$ \Rvir\ ($\sim$500\,pc) for the m12i simulation (Figure~\ref{fig:m12i_turbsig}). The results are shown in Figure~\ref{fig:m12i_turbsig}.
In general, snapshots with larger \sigoned\ tend to have larger \sigintr\, and vice versa.
The offset at lower redshift may be due to \sigintr\ including some differential rotation on kiloparsec scales, whereas \sigoned\ minimises this effect by construction.
This suggests that the quantity defined to mimic intrinsic velocity dispersion measurements, \sigoned, does broadly reflect the local random motion of the overall gas content, as probed by \sigintr, with a caveat of large scatter.

\subsection{Observational implications and future work}

In this section, we discuss several directions for future work.
First, whereas in this work, we derive a single SFR-weighted $\sigma$ for individual galaxies when analysing the simulations and comparing them with observations, we can extend this analysis to spatially resolved $\sigma$, SFR, and $M_{\rm gas}$ maps and then compare the simulations with a subset of AO-assisted IFS observations with kpc-scale resolution.
Such high-resolution observations have revealed $\sigma$ maps that are far from spatially uniform, and the highest $\sigma$ values can occur at the outskirts of galaxies (and are thus unlikely to be a result of beam-smearing
effects) and may or may not correspond to local maxima in the SFR surface density \citep[e.g.][]{Swinbank2012,Livermore2015}.
This spatial non-uniformity is perhaps not surprising because the dominant scale on which stellar feedback drives turbulence is of order the disc scaleheight \citep{Hayward2017} and
thus can be $\lesssim1$\,kpc. Moreover, different physical drivers may be relevant on different scales or/and in different physical regimes (e.g. at different gas surface densities).
Searching for correlations between $\sigma$ and other parameters at kiloparsec scales may therefore be useful for determining the physical driver(s) of local velocity dispersion enhancements,
although the timescale effects explored above may hamper the physical interpretation of such spatially resolved observations.

Our analysis demonstrates the importance of gas accretion in driving enhancements in velocity dispersion -- perhaps indirectly by driving enhancements in the SFR and thus stellar feedback --
but it is extremely challenging to directly probe the relationship between inflow and galaxy kinematics observationally.
Although current observational constraints on \Minflow\ are limited to only a few candidate galaxies, examples such as that presented by \citet{Bouche2013} are intriguing given that their \Minflow\ estimate of $\sim$30-60 \Msun\,yr$^{-1}$ at 26 kpc from the galaxy centre is comparable to the inflow rates of most of the simulated galaxies studied in this work, and such \Minflow\ rates are indeed reasonable for galaxies with \sigoned\ $\gtrsim50$ km\,s$^{-1}$.

Moreover, individual galaxies in the FIRE simulations experience multiple phases with no ongoing star formation (although some low level of star formation may occur if the galaxies were simulated at higher resolution)
with untraceable \sigoned.
These galaxies can still have high $f_{\rm total gas}$ values (based on the total gas mass in the simulations) and are only temporarily quenched and likely to rejuvenate within a few hundred Myr. 
If such periods are indeed important evolutionary phases of galaxies, then there would be a population of high-$f_{\rm total gas}$ galaxies with low SFR that would not be included in most IFS surveys because they would be undetected or have too low signal-to-noise ratios for detailed kinematic analysis.
Future IFS surveys that can probe stellar kinematics of high$-z$ galaxies are thus critical for probing the kinematic evolution of galaxies in all evolutionary phases.

\section{Conclusions}
We investigate the origin of enhanced gas velocity dispersion (\sigintr) of high-redshift star-forming galaxies using a suite of cosmological simulations from the FIRE-1 project.  
We define a measure of \sigoned\, the minimum SFR-weighted standard deviation computed over 10$^4$ viewing angles, that aims to be representative of \sigintr\ for observed galaxies.
In parallel, we compile a set of \sigintr\ measurements of $0\leq z\lesssim3$ star-forming galaxies from IFS surveys that probe the kinematics of ionised gas, and  we systematically compare theses observations
with the simulated galaxies from the FIRE-1 project.
Our primary conclusions are summarised as follows:
  
\begin{itemize}
\item The 1-D velocity dispersions of the galaxies in the FIRE simulations exhibit a trend with redshift similar to that observed, i.e. the median \sigoned\ increases from $z\sim0$ to $z\sim1$ and flattens beyond $z\gtrsim1$.
The trend with redshift exhibited by both the simulations and the observations compiled in this work is inconsistent with the monotonic increase found by \citet{Wisnioski2015}; for the observations, the discrepancy between
their results and ours is primarily a result of the $z\sim1$ KMOS$^{\rm 3D}$ sample analysed by \citet{Wisnioski2015} having velocity dispersions that are systematically $\sim20-30$ km\,s$^{-1}$ less than found in other surveys.

\item The simulated galaxies exhibit a positive correlation between \sigoned\ and SFR, as is also evident in the observations.
In both the observed and simulated galaxy samples, there is no obvious offset in the \sigoned--SFR relations at different redshifts.

\item The large scatter in the \sigoned--SFR relation suggests that either stellar feedback is not the only relevant driver of enhanced velocity dispersion or that the slight asynchronicity of the enhancements
in \sigoned and SFR results in a large physical scatter in this relation.

\item A possible correlation of \sigoned\ with \Mstar, which is seen in the KROSS galaxies at $z\sim1$ and the FIRE simulations at $z\gtrsim2$, suggests that differences in \Mstar\ at a fixed SFR may contribute to the
scatter in the \sigoned--SFR relation.

\item Based on an analysis of the time evolution of individual simulated galaxies' physical and kinematic properties (SFR, \Mstar, \Minflow, $f_{\rm gas}$ and \sigoned), we find that the variations in SFR,
\Minflow, and $f_{\rm gas}$ are approximately temporally coincident with variations in \sigoned, whereas \Mstar\ varies on longer timescales.
SFR, \Minflow, and \sigoned\ can all vary significantly on timescales of $<100$ Myr, especially at high redshift ($z\gtrsim1$).

\item We measure the time delay between pairs of the three properties SFR, \Minflow, and \sigoned\ by cross-correlating their time series.
A robust negative time delay between \Minflow\ and SFR is found (i.e. peaks in \Minflow\ tend to proceed peaks in SFR), thus suggesting that variations in the gas inflow rate through a shell of radius 0.2 \Rvir,
due either to cosmological inflow or galactic fountains, lead to subsequently enhanced SFR.
The small time delays measured between \Minflow\ and \sigoned\ and between SFR and \sigoned, the signs of which can depend on the simulation and details of the cross-correlation measurement,
indicate that the variations in velocity dispersion temporally coincide with variations in SFR and \Minflow, but the causality (i.e. whether accretion or/and stellar feedback drives enhancements
in \sigoned) cannot be determined from this analysis.
\end{itemize}

Future work comparing AO-assisted IFS observations with spatially resolved properties of simulated galaxies should yield further insights into what physical processes drive turbulence in
galaxies.

\section*{Acknowledgements}
We thank J. Stott and the KROSS team for kindly sharing their measurements of galaxy properties.
C-LH acknowledges support from the Harlan J. Smith Fellowship at the University of Texas at Austin.
The Flatiron Institute is supported by the Simons Foundation. 
TY acknowledges fellowship support from the Australian Research Council Centre of Excellence for All Sky Astrophysics in 3 Dimensions (ASTRO 3D), through project number CE170100013.
CAFG was supported by NSF through grants AST-1412836, AST-1517491, AST-1715216, and CAREER award AST-1652522, by NASA through grant NNX15AB22G, and by a Cottrell Scholar Award from the Research Corporation for Science Advancement.
Support for PFH was provided by an Alfred P. Sloan Research Fellowship, NASA ATP Grant NNX14AH35G, and NSF Collaborative Research Grant \#1411920 and CAREER grant \#1455342. 
DK was supported by NSF grant AST-1715101 and the Cottrell Scholar Award from the Research Corporation for Science Advancement.
AW was supported by NASA through grants HST-GO-14734 and HST-AR-15057 from STScI.

The numerical calculations were run on the Caltech compute cluster ``Zwicky'' (NSF MRI award \#PHY-0960291), allocations TG-AST120025 and TG-AST130039 granted by the Extreme Science and Engineering Discovery Environment (XSEDE) supported by the NSF, and allocation PRAC NSF.1713353 supported by the NSF.
The authors acknowledge the Texas Advanced Computing Center (TACC) at The University of Texas at Austin for providing HPC resources that have contributed to the research results reported within this paper. URL: \href{http://www.tacc.utexas.edu}{http://www.tacc.utexas.edu}




\bibliographystyle{mnras}
\bibliography{fire,cosmos}






\bsp	
\label{lastpage}
\end{document}